\documentclass[
  aps,
  physrev,
  reprint,
  floatfix,
  superscriptaddress,
  amsmath,
  amssymb,
  longbibliography
]{revtex4-2}

\usepackage{graphicx}
\usepackage{bm}
\usepackage{hyperref}
\usepackage{microtype}
\usepackage{placeins}
\hypersetup{colorlinks=true,linkcolor=blue,citecolor=blue,urlcolor=blue}

\begin{document}

\title{Finite-size scaling and cutoff functions of critical avalanches under cubic saturation}

\author{Ryota Kojima}
\email{ryota_kojima_aa@mail.toyota.co.jp}
\affiliation{Frontier Research Center, Toyota Motor Corporation, Aichi, Japan}

\author{Kazufumi Hosoda}
\affiliation{Center for Information and Neural Networks, Advanced ICT Research Institute, National Institute of Information and Communications Technology, Osaka, Japan}
\affiliation{Life and Medical Sciences Area, Health Sciences Discipline, Kobe University, Hyogo, Japan}

\author{Hiroyuki Suto}
\affiliation{Frontier Research Center, Toyota Motor Corporation, Aichi, Japan}

\author{Hitoshi Yamada}
\affiliation{Frontier Research Center, Toyota Motor Corporation, Aichi, Japan}

\begin{abstract}
Power-law avalanche statistics characterize the branching bulk but leave open how finite systems terminate their largest events. Here we determine both the finite-size cutoff and the full crossover function connecting the branching regime to the large-event tail for an exchangeable absorbing event process motivated by Cauchy-stable recurrent dynamics. The Cauchy dynamical mean-field theory yields a linear-cubic low-activity drift, while a repeatable-event lift supplies absorbing branching fluctuations. Their competition at criticality generates a distinct finite-size edge controlled by cubic saturation. For avalanches initiated by a fixed microscopic seed, the rescaled edge measure connects the universal branching power law continuously to an exponential large-area tail. This full crossover admits an analytic spectral representation in terms of the boundary spectrum of a quartic Schrödinger operator on the half-line: high modes reconstruct the branching singularity, while the ground state fixes the far-tail decay. Exact finite-size calculations and Monte Carlo simulations confirm both the predicted cutoff behavior and the full spectral crossover.
\end{abstract}
\maketitle

\section{Introduction}
\label{sec:introduction}

Neuronal population activity often appears in temporally localized bursts, or avalanches, whose sizes and durations span broad ranges
\cite{Beggs11167,PlenzThiagarajan2007,Petermann2009,
PhysRevLett.108.208102,Klaus2011,Shriki7079,hahn2010neuronal}.
Their approximate power-law statistics have motivated the proposal that cortical circuits operate near a critical regime, with possible consequences for dynamic range, information transmission, and flexible collective responses
\cite{KinouchiCopelli2006,Shew15595,ShewPlenz2013,Wilting201925}.
Bulk exponents alone, however, do not specify how the largest avalanches are terminated. Universality can also reside in scaling functions and avalanche shapes \cite{sethna2001crackling,papanikolaou2011universality}, while experimentally inferred cutoffs can depend on the observation window and measurement protocol \cite{Yu2014}. Power-law scaling is therefore insufficient by itself to identify either criticality or the mechanism that suppresses large events \cite{touboul2017power}.

Critical branching processes and Feller diffusions with square-root noise provide the standard mean-field description of the avalanche bulk
\cite{Harris1963,AthreyaNey1972,PhysRevLett.94.058101,diSanto2017,garcia2018field,le2022equivalence}.
Finite-size survival has been studied in explicitly size-limited branching processes, and extinction, quasi-stationarity, transition laws, and killed spectra have been analyzed for generalized and logistic Feller diffusions
\cite{GarciaMillan2015,Cattiaux2009,gan2015singular,foucart2025conditioning}.
Exact distributions are also available for several mean-field avalanche models \cite{delorme2016distribution}.
These results establish much of the branching bulk and long-time absorbing dynamics, but do not by themselves determine how a specified nonlinear saturation term fixes both the finite-size cutoff scale and the full crossover function for avalanches initiated by a microscopic seed. Here we address this problem for leading cubic saturation.

A linear-cubic absorbing normal form with square-root noise also appears in mean-field descriptions of tricritical directed percolation, where the quadratic nonlinear coupling vanishes and cubic saturation becomes leading \cite{ohtsuki1987nonequilibrium,lubeck2006tricritical}. Related tricritical absorbing behavior has also been reported in stochastic Wilson--Cowan neural dynamics \cite{piuvezam2023unconventional}. In the Cauchy--Heaviside model considered here, the absence of the quadratic term follows structurally from the threshold response. We investigate the resulting finite-size avalanche statistics, focusing on the cutoff generated from a fixed microscopic seed.

A tractable realization of cubic saturation arises in a recurrent rate network with Cauchy-distributed couplings. Classical random-network theories usually assume light-tailed connectivity
\cite{sompolinsky1988chaos,rajan2010stimulus}, whereas measured synaptic strengths are strongly heterogeneous
\cite{Sayer826,https://doi.org/10.1111/j.1469-7793.1999.00169.x,
song2005highly,Buzsaki2014,Cossell2015}.
Heavy-tailed recurrent connectivity can modify the instability transition, generate scale-free avalanches, and produce extended critical regimes
\cite{PhysRevLett.125.028101,PhysRevResearch.3.013083,wardak2022extended}.
The Cauchy case is analytically distinguished by stability under summation: for nonnegative activity, the recurrent input is described by a single scale parameter. Near the transition from quiescence to self-sustained activity, the associated dynamical mean-field theory (DMFT) reduces to linear growth stabilized by a cubic term
\cite{kojima2026robustnearcriticaldynamicsheavytailed}.

\begin{figure*}[t]
  \centering
  \includegraphics[width=\textwidth]{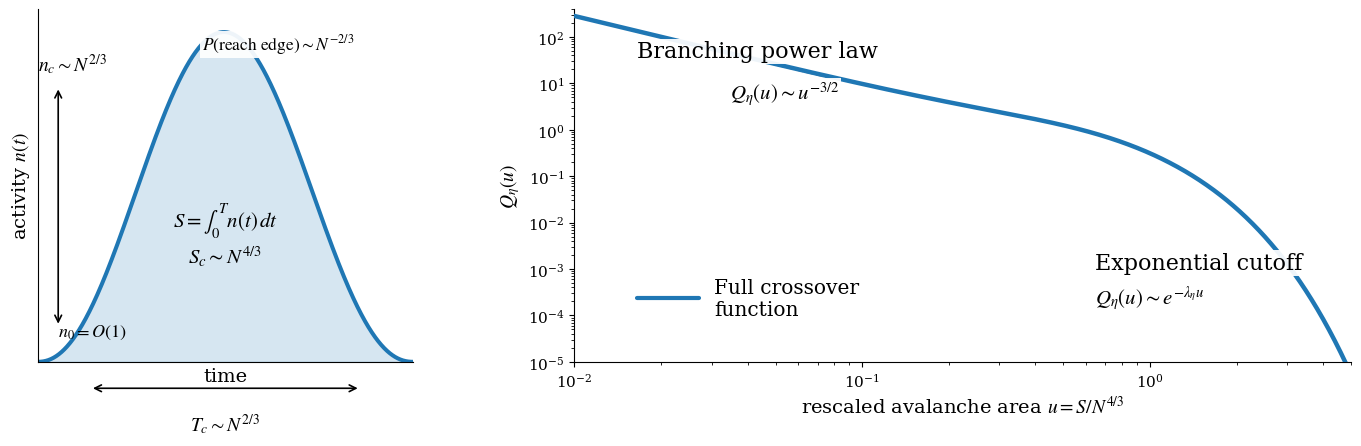}
  \caption{\textbf{Finite-size avalanche scales and crossover.}
  Left: Schematic of a rare avalanche reaching the cubic-saturation edge. The characteristic active-event count, duration, and area scales are $n_c,T_c\sim N^{2/3}$ and $S_c\sim N^{4/3}$, with an $O(N^{-2/3})$ probability of reaching this regime from a fixed microscopic seed.
  Right: Full crossover of the limiting rescaled area density $Q_\eta(u)$, where $u=S/N^{4/3}$. For the fixed seed $n_0=1$, $Q_\eta$ denotes the non-normalized limiting density obtained by rescaling rare-event probabilities by $N^{2/3}$. The curve is calculated from the quartic boundary spectrum at criticality ($\eta=0$) and connects the branching power law $u^{-3/2}$ to an exponential tail with decay rate $\lambda_\eta$. Here $\eta$ measures the dimensionless detuning from criticality.}
  \label{fig:overview-main}
\end{figure*}

The deterministic DMFT captures the collective drift but does not determine the fluctuations of individual avalanches. To study their finite-size statistics, we introduce an exchangeable birth--death process in which each of the $N$ network units can repeatedly generate events. The number $n(t)$ counts all events active at time $t$, including multiple events from the same unit. An avalanche begins with a fixed microscopic seed $n(0)=n_0=O(1)$ and ends upon extinction at time $T$. Its area is defined as $S=\int_0^T n(t)\,dt$. The birth and death rates are chosen so that the hydrodynamic drift reproduces the Cauchy DMFT, while the fluctuations have the square-root form characteristic of absorbing branching processes. This construction is a mesoscopic realization of the mean-field mode, not a pathwise reduction of a fixed quenched network.

Our main results are summarized in Fig.~\ref{fig:overview-main}. Cubic saturation leaves the small-avalanche branching regime intact but controls rare large excursions. The characteristic activity and duration scales grow as $n_c,T_c\sim N^{2/3}$, giving an avalanche-area cutoff $S_c\sim N^{4/3}$ (left panel). A fixed microscopic seed reaches this edge with probability $O(N^{-2/3})$. This delayed finite-size arrest distinguishes cubic saturation from leading quadratic saturation. More importantly, we determine not only these cutoff scales but also the full crossover from the branching power law to the exponential large-area tail (right panel).

To obtain the full crossover function, we analyze the exact finite-$N$ backward equation for the area transform generated from a fixed microscopic seed. In the critical window, a discrete boundary-matching argument reduces the fixed-seed transform to the boundary response of a universal Feynman--Kac problem. A gauge transformation then maps that problem to a quartic Schr\"odinger operator on the half-line. Its Dirichlet eigenvalues determine the decay scales, while the squared boundary derivatives of the eigenfunctions determine the spectral weights. High modes reconstruct the branching singularity, whereas the ground state fixes the exponential far tail. Unlike a macroscopic-initial-condition spectral expansion, the microscopic seed enters through boundary flux rather than a bulk overlap.

We test the theory using exact finite-$N$ moment and backward equations, generalized area-tail eigenvalues, and Monte Carlo simulations of the exact exchangeable Cauchy--Heaviside population rates. These calculations separately probe the cutoff exponents, convergence to the microscopic-seed boundary law, and the quartic spectral tail.

Section~\ref{sec:dmft} reviews the Cauchy DMFT and extracts its low-activity normal form. Section~\ref{sec:event-process} defines the exchangeable finite-$N$ process and its edge generator. Section~\ref{sec:bulk-edge} establishes the branching bulk and the soft-edge scales. Section~\ref{sec:spectral} derives the boundary spectral law, and Sec.~\ref{sec:numerical-main} presents the finite-size tests. The Supplemental Material contains the microscopic-seed matching proof, further derivations, and details of the numerical methods. Principal notation is summarized in Table~\ref{tab:notation}.

\begin{table*}[t]
\caption{\textbf{Principal notation.} Quantities used across more than one stage of the construction; local auxiliary symbols are defined only where they are needed.}
\label{tab:notation}
\begin{ruledtabular}
\begin{tabular}{clcl}
Symbol & Meaning & Symbol & Meaning\\
\hline
$N$ & number of network units/source labels & $\tau_0$ & microscopic network and event timescale\\
$g$ & Cauchy coupling scale & $\Delta=\theta-h$ & threshold gap\\
$\sigma$ & recurrent-input scale in the Cauchy DMFT & $n(t)$ & number of active events\\
$W^\pm(n)$ & population birth/death rates & $A_N,B_N$ & first and second jump-cumulant rates\\
$\kappa$ & finite-size critical-window detuning & $\beta$ & cubic saturation coefficient\\
$\Gamma$ & square-root-noise amplitude & $D=\Gamma^2/2$ & edge diffusion coefficient\\
$S,T$ & avalanche area and duration & $n_0$ & microscopic seed size\\
$X_s$ & rescaled edge activity & $\ell$ & natural edge activity scale\\
$\eta$ & dimensionless detuning & $\widehat\lambda,\Lambda$ & edge-scale and dimensionless area-Laplace variables\\
$E_j,\rho_j$ & spectral decay rates and boundary weights & $G_\eta$ & dimensionless cutoff function\\
\end{tabular}
\end{ruledtabular}
\end{table*}

\section{Rate-network origin of the cubic drift}
\label{sec:dmft}

The deterministic Cauchy DMFT derived in Ref.~\cite{kojima2026robustnearcriticaldynamicsheavytailed} fixes the low-activity drift that the finite-size event process must reproduce.
We summarize only the ingredients needed here; finite-$N$ fluctuations are introduced in Sec.~\ref{sec:event-process}.

\subsection{Cauchy recurrent network}

We consider
\begin{equation}
 \tau_0\dot x_i(t)
 =-x_i(t)+\sum_{j=1}^N J_{ij}\phi(x_j(t))+h_i(t),
 \label{eq:rate-network}
\end{equation}
for $i=1,\ldots,N$, where $\tau_0>0$, $h_i(t)$ is an external drive, and the nonlinearity is bounded, nonnegative, and nondecreasing,
$ 0\leq \phi(x)\leq \phi_{\max}, $
while the couplings are independent symmetric Cauchy variables,
\begin{equation}
 \left\langle e^{ikJ_{ij}}\right\rangle
 =
 \exp\left(-\frac{g}{N}|k|\right).
 \label{eq:cauchy-couplings}
\end{equation}
The scale $g/N$ is the stable-law analogue of mean-field normalization: a sum of $O(N)$ couplings remains $O(1)$.

For a constant homogeneous drive $h_i(t)=h$, take
$ x_i(0)=h+\xi_i,$ where the $\xi_i$ are independent centered Cauchy variables of scale $\sigma_0\geq0$, independent of the couplings.
Define the filtered activity of source $j$ by
\begin{equation}
 a_j(t)
 =
 \frac1{\tau_0}
 \int_0^t e^{-(t-s)/\tau_0}\phi(x_j(s)) ds.
 \label{eq:filtered-activity}
\end{equation}
Solving Eq.~\eqref{eq:rate-network} gives
$ x_i(t)-h
 =
 e^{-t/\tau_0}\xi_i
 +
 \sum_{j=1}^N J_{ij}a_j(t).$
Because $\phi\geq0$, one has $a_j(t)\geq0$.
Conditional on the filtered activities, Cauchy stability therefore gives a centered Cauchy fluctuation about $h$ with scale
$ \sigma_0 e^{-t/\tau_0}
 +
 \frac{g}{N}\sum_{j=1}^N a_j(t).$
In the exchangeable large-$N$ limit, define the mean activity
$ \bar m(t)=\langle\phi(x(t))\rangle, $ where $\langle\cdot\rangle$ denotes the single-site ensemble average.
The Cauchy scale then self-averages to
\begin{equation}
 \sigma(t)
 =
 \sigma_0 e^{-t/\tau_0}
 +
 \frac{g}{\tau_0}
 \int_0^t e^{-(t-s)/\tau_0}\bar m(s) ds.
 \label{eq:sigma-filter}
\end{equation}

\subsection{Linear-cubic low-activity mode}

We now specialize to the threshold nonlinearity $\phi(x)=\Theta(x-\theta)$, where $\Theta$ is the Heaviside step function and $\theta$ is the activation threshold, and to a homogeneous subthreshold drive $h<\theta$.
Define $ \Delta=\theta-h>0.$
Let $Z_\sigma$ denote a centered Cauchy variable of scale $\sigma$.
The Cauchy-smoothed response is
\begin{align}
 m_\Delta(\sigma)
 &:=\mathbb P(Z_\sigma>\Delta)\nonumber\\
 &=\frac{1}{2}-\frac{1}{\pi}\arctan\frac{\Delta}{\sigma}
 =\frac{1}{\pi}\arctan\frac{\sigma}{\Delta}.
 \label{eq:cauchy-response}
\end{align}
For the homogeneous threshold model, $\bar m(t)=m_\Delta(\sigma(t))$. Differentiating Eq.~\eqref{eq:sigma-filter} gives
\begin{equation}
 \tau_0\dot\sigma
 =
 -\sigma+g m_\Delta(\sigma),
 \quad
 \sigma(0)=\sigma_0.
 \label{eq:dmft-ode}
\end{equation}
Equation~\eqref{eq:dmft-ode} is the deterministic large-$N$ DMFT equation derived in Ref.~\cite{kojima2026robustnearcriticaldynamicsheavytailed}.

The small-$\sigma$ expansion
\begin{equation}
 m_\Delta(\sigma)
 =
 \frac{\sigma}{\pi\Delta}
 -\frac{\sigma^3}{3\pi\Delta^3}
 +O(\sigma^5)
 \label{eq:cauchy-response-expansion}
\end{equation}
gives the low-activity normal form
\begin{equation}
 \tau_0\dot\sigma
 =
 \left(\frac{g}{\pi\Delta}-1\right)\sigma
 -\frac{g}{3\pi\Delta^3}\sigma^3
 +O(\sigma^5).
 \label{eq:dmft-normal-form}
\end{equation}
The quiescent solution is $\sigma=0$.
Since $m_\Delta'(0)=1/(\pi\Delta)$, its linear stability changes when $g m_\Delta'(0)=1$. 
The critical coupling is therefore
\begin{equation}
 g_c=\pi\Delta.
 \label{eq:critical-coupling}
\end{equation}
At $g=g_c$, the leading deterministic decay is algebraic,
$
 \sigma(t)
 \sim
 \left(\frac{2t}{3\Delta^2\tau_0}\right)^{-1/2}.
$

The missing quadratic term is not implied by Cauchy stability alone; it follows from the odd small-$\sigma$ expansion of the Heaviside-threshold response in Eq.~\eqref{eq:cauchy-response}. 
This is the structural feature that selects the cubic finite-size edge.
A transfer function or finite-size lift that generates a quadratic correction instead enters a different edge class, as shown in Sec.~\ref{sec:event-process}.

\section{Finite-\texorpdfstring{$N$}{N} event process and edge generator}
\label{sec:event-process}

The DMFT fixes the deterministic drift but not the finite-$N$ fluctuations that generate avalanches. We therefore define a concrete exchangeable birth--death lift and determine both its hydrodynamic limit and the jump cumulants that survive at the critical edge. The finite-size behavior also depends on how the event process is constructed.

\subsection{Repeatable-event construction}

At time $t$, let
$
 \mathcal P(t)=\{j_1,\ldots,j_{n(t)}\}
$
be a multiset of active source labels; repeated labels represent distinct events emitted by the same source. For a fixed coupling realization $J$, the instantaneous input to target $i$ is
\begin{equation}
 I_i(\mathcal P)=h+\sum_{r=1}^{n}J_{ij_r}.
 \label{eq:event-input}
\end{equation}
Each active event has an independent exponential lifetime of mean $\tau_0$, giving
\begin{equation}
 W^-(n)=\frac{n}{\tau_0}.
 \label{eq:death-rate}
\end{equation}
Each target whose instantaneous input exceeds the threshold, $I_i(\mathcal P)>\theta$, emits a new event at rate $1/\tau_0$. The resulting fixed-disorder birth rate is
\begin{equation}
 W_J^+(\mathcal P)=\frac1{\tau_0}\sum_{i=1}^N
 \Theta\bigl(I_i(\mathcal P)-\theta\bigr).
 \label{eq:quenched-birth}
\end{equation}

We define the exchangeable population generator by averaging Eq.~\eqref{eq:quenched-birth} over the Cauchy ensemble and over exchangeable label multisets of cardinality $n$:
\begin{equation}
 \overline W^+(n)=
 \mathbb E_J\mathbb E_{\mathcal P\mid |\mathcal P|=n}
 \left[W_J^+(\mathcal P)\right].
 \label{eq:averaged-rate-def}
\end{equation}
The resulting rate does not depend on the particular exchangeable distribution over such multisets: all multiplicities are nonnegative and sum to $n$. Counting repeated labels with their multiplicity, Cauchy stability gives
$
 \sum_{r=1}^{n}J_{ij_r}\sim
 \mathrm{Cauchy}\left(0,gn/N\right).
$
Hence
\begin{equation}
 \overline W^+(n)=\frac{N}{\tau_0}
 m_\Delta\left(\frac{gn}{N}\right).
 \label{eq:population-birth}
\end{equation}
We henceforth denote this exchangeable population rate by $W^+(n)$. Together, Eqs.~\eqref{eq:death-rate} and \eqref{eq:population-birth} define a closed absorbing birth--death process.

This ensemble average is not a quenched closure. For fixed $J$, the label process $\mathcal P(t)$ is Markovian, but its cardinality $n(t)$ need not be, because dynamically selected labels can correlate with atypical coupling columns. All results below refer to the exchangeable population process defined by Eq.~\eqref{eq:population-birth}.

\subsection{Hydrodynamic match and finite-capacity correction}

Define $\sigma_N=gn/N$. A jump changes $\sigma_N$ by $\pm g/N$, and for the repeatable-event process its drift is
\begin{equation}
 \frac{g}{N}\left[W^+(n)-W^-(n)\right]
 =\frac1{\tau_0}\left[g m_\Delta(\sigma_N)-\sigma_N\right].
 \label{eq:hydrodynamic-match}
\end{equation}
For macroscopic initial conditions, the accumulated variance of $\sigma_N$ is $O(N^{-1})$, so the density-dependent fluid limit is Eq.~\eqref{eq:dmft-ode} \cite{kurtz1970solutions}.

A finite-capacity constraint changes this structure. If emission is restricted to inactive sites, the birth rate becomes
\begin{equation}
 W^+_{\mathrm{occ}}(n)=\frac{N-n}{\tau_0}
 m_\Delta\left(\frac{gn}{N}\right).
 \label{eq:occupancy-birth}
\end{equation}
Its low-activity drift is
\begin{align}
 W^+_{\mathrm{occ}}(n)-W^-(n)
 =\frac1{\tau_0}\biggl[
 &\left(\frac{g}{\pi\Delta}-1\right)n
 -\frac{g}{\pi\Delta}\frac{n^2}{N}
 \nonumber\\
 &-\frac{g^3}{3\pi\Delta^3}\frac{n^3}{N^2}
 +\cdots
 \biggr].
 \label{eq:occupancy-drift}
\end{align}
The factor $1-n/N$ therefore generates a quadratic correction before the cubic term.
At criticality, this term becomes relevant at $n\sim N^{1/2}$, whereas the cubic correction becomes relevant only at $n\sim N^{2/3}$.
For $n=O(N)$, the same finite-capacity factor also contributes to the macroscopic drift.

\subsection{Jump cumulants and edge generator}

For the Cauchy--Heaviside process,
\begin{align}
 W^+(n)=\frac1{\tau_0}\biggl[&\frac{g}{\pi\Delta}n
 -\frac{g^3}{3\pi\Delta^3}\frac{n^3}{N^2}\nonumber\\
 &+\frac{g^5}{5\pi\Delta^5}\frac{n^5}{N^4}
 +O\left(\frac{n^7}{N^6}\right)\biggr].
 \label{eq:Wexpansion}
\end{align}
The drift and quadratic-variation rates are
\begin{align*}
 A_N(n)&=W^+(n)-W^-(n),\\
 B_N(n)&=W^+(n)+W^-(n).
\end{align*}
Anticipating the soft-edge balance derived in Sec.~\ref{sec:bulk-edge}, we allow the coupling scale to depend on system size, $g=g_N$, and tune it within the finite-size critical window as
\begin{equation}
 \frac{g_N}{\pi\Delta}-1=\kappa N^{-2/3},
 \label{eq:critical-window-tuning}
\end{equation}
with fixed $\kappa=O(1)$. On the edge scale, the two cumulants take the form
\begin{align}
 A_N(n)&=\frac1{\tau_0}\left(
 \kappa N^{-2/3}n-\beta\frac{n^3}{N^2}\right)
 +o\left(\frac{n^3}{N^2}\right),\nonumber\\
 B_N(n)&=\Gamma^2 n+o(n).
 \label{eq:cumulant-class}
\end{align}
For the Cauchy--Heaviside model at criticality,
\begin{equation}
 \beta=\frac{\pi^2}{3},\quad
 \Gamma^2=\frac{2}{\tau_0},\quad
 D=\frac{\Gamma^2}{2}.
 \label{eq:critical-coefficients}
\end{equation}
Retaining these two cumulants gives the mesoscopic diffusion approximation
\begin{equation}
 dn_t=
 \frac{1}{\tau_0}\left(
 \kappa N^{-2/3}n_t-\beta\frac{n_t^3}{N^2}\right)dt
 +\Gamma\sqrt{n_t} d\mathcal B_t,
 \label{eq:mesoscopic-sde}
\end{equation}
where $\mathcal B_t$ is a standard Brownian motion and $n=0$ is absorbing. Equation~\eqref{eq:mesoscopic-sde} displays the competing drift and fluctuation scales, but it is not taken as the microscopic model. The limiting diffusion is derived below from the exact birth--death generator.

The arctangent response is one representative of the cubic class. More generally, the same edge structure follows when the low-activity response is $c_1\sigma-c_3\sigma^3+o(\sigma^3)$ with $c_1,c_3>0$, after tuning $gc_1=1$. By contrast, a quadratic response term produces a drift of order $n^2/N$. The same correction is generated by the finite-capacity factor $1-n/N$ in Eq.~\eqref{eq:occupancy-birth}. It becomes relevant at $n\sim N^{1/2}$, before the cubic scale $n\sim N^{2/3}$, and therefore defines a distinct finite-size class.

The resulting edge limit is not specific to the arctangent rate. It depends only on the rescaled jump statistics. Specifically, the limits in Eq.~\eqref{eq:cumulant-class} must hold uniformly on every compact interval $x\in[\varepsilon,L]\subset(0,\infty)$, with $n=N^{2/3}x$, while all rescaled jump cumulants of order three and higher must vanish. Under these conditions, different microscopic processes converge to the same edge diffusion. The spectral law for a fixed microscopic seed additionally requires the near-boundary conditions stated in the Supplemental Material.

The exact discrete generator is
\begin{multline}
 \mathcal G_N f(n)=W^+(n)[f(n+1)-f(n)]\\
 +W^-(n)[f(n-1)-f(n)].
 \label{eq:discrete-generator}
\end{multline}
Define the rescaled process
\begin{equation}
 X_s^{(N)}=N^{-2/3}n_{N^{2/3}s}.
 \label{eq:rescaled-process}
\end{equation}
For a smooth test function $\varphi$, let $f_N(n)=\varphi(n/N^{2/3})$. A Taylor expansion of the exact generator gives, uniformly on fixed intervals $x\in[\varepsilon,L]$,
\begin{align}
 N^{2/3}\mathcal G_N f_N(n)
 &=
 \frac{\kappa x-\beta x^3}{\tau_0}\varphi'(x)
 +Dx\varphi''(x)\nonumber\\
 &\quad+O(N^{-2/3}),
 \quad x=\frac{n}{N^{2/3}}.
 \label{eq:generator-convergence}
\end{align}
Equation~\eqref{eq:generator-convergence} gives the diffusion limit of the rescaled density-dependent jump process \cite{kurtz1971limit}. The limiting absorbing diffusion is
\begin{equation}
 dX_s=\frac{\kappa X_s-\beta X_s^3}{\tau_0} ds
 +\Gamma\sqrt{X_s} d\mathcal B_s.
 \label{eq:edge-diffusion}
\end{equation}
Thus the mesoscopic SDE in Eq.~\eqref{eq:mesoscopic-sde} is recovered from the discrete process rather than postulated independently.

\begin{figure*}[t]
 \centering
 \includegraphics[width=\textwidth]{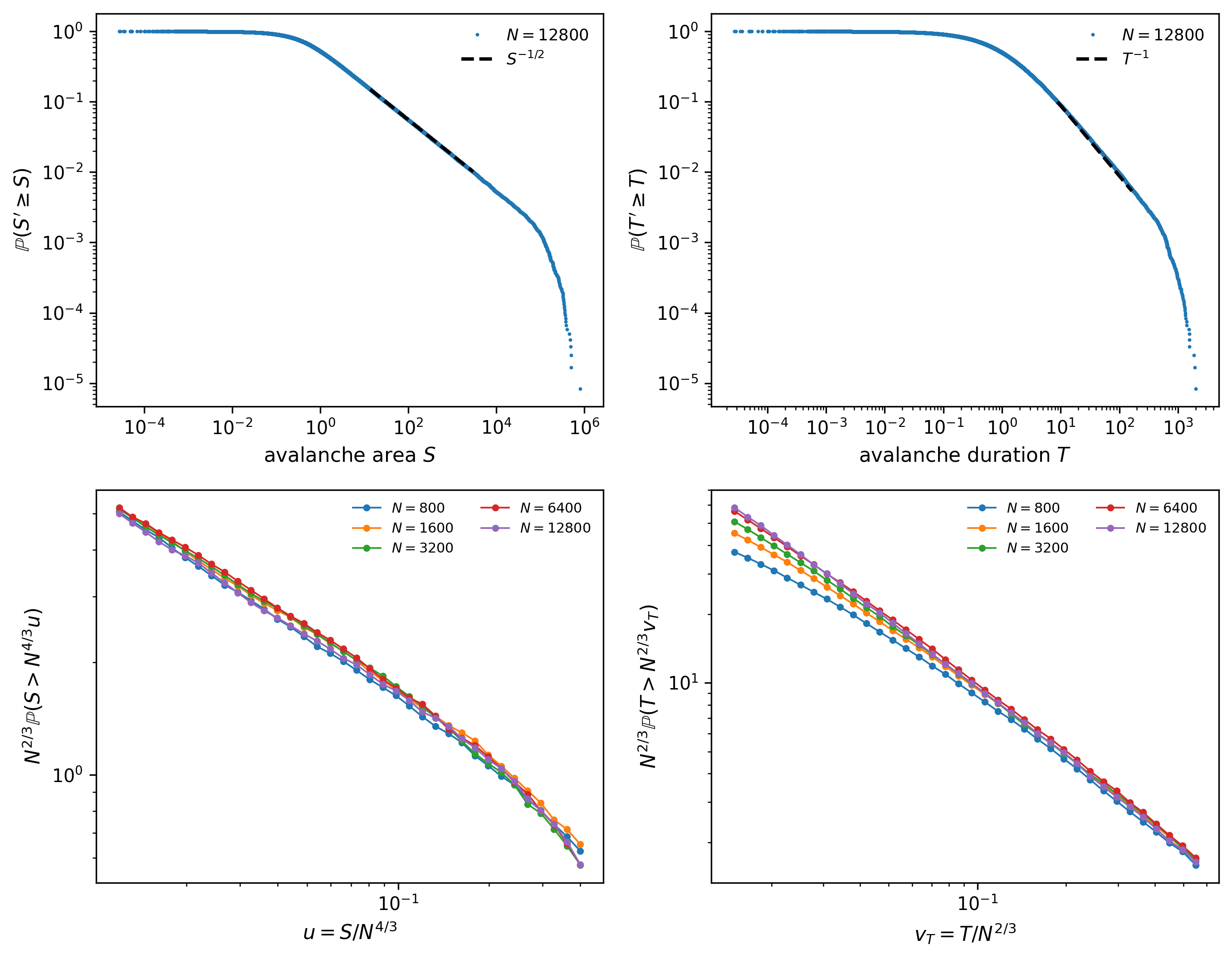}
\caption{\textbf{Branching bulk and finite-size edge scaling.}
Upper panels show complementary cumulative distributions (CCDFs) of avalanche area $S$ (left) and duration $T$ (right) at $N=12800$ for the critical exchangeable Cauchy--Heaviside process.  The reference laws $S^{-1/2}$ and $T^{-1}$ are the cumulative counterparts of the bulk density laws $p_S(S)\sim S^{-3/2}$ and $p_T(T)\sim T^{-2}$ in Eqs.~\eqref{eq:bulk-laws} and \eqref{eq:duration-bulk}. Lower panels show the finite-size edge collapse of the area tail under $S/N^{4/3}$ (left) and of the duration tail under $T/N^{2/3}$ (right), consistent with Eq.~\eqref{eq:cutoff-scales}.  In both panels the probability weight is multiplied by $N^{2/3}$, reflecting the $O(N^{-2/3})$ probability for a fixed microscopic seed to reach the edge scale [cf.~Eq.~\eqref{eq:edge-measure}].
The five system sizes are $N=800,1600,3200,6400,12800$. All critical simulations use $\Delta=\tau_0=1$, $n_0=1$, and $g_N=g_c=\pi$; trajectory counts for each $N$ are listed in Table~\ref{tab:parameters}.}
 \label{fig:bulk-edge-main}
\end{figure*}
\section{Branching bulk and the cubic soft edge}
\label{sec:bulk-edge}

The critical process separates into a branching bulk and a saturation-controlled edge. Far below the edge, the cubic drift is asymptotically negligible; near the edge, it competes with branching fluctuations. We first recover the bulk area and duration laws and then determine the activity, time, and area scales of the crossover. Section~\ref{sec:spectral} derives its shape.

\subsection{Squared-Bessel branching bulk}

At criticality and below the edge scale, the cubic drift is negligible and
\begin{equation}
 dn(t)=\Gamma\sqrt{n(t)} d\mathcal B_t.
 \label{eq:critical-feller}
\end{equation}
The Lamperti transform $Y_t=2\sqrt{n(t)}/\Gamma$ gives
\begin{equation}
 dY_t=d\mathcal B_t-\frac{1}{2Y_t} dt,
\end{equation}
so $4n(t)/\Gamma^2$ is a squared-Bessel process of dimension zero (BESQ$^{0}$) \cite{feller1951diffusion,lamperti1967limit,revuz2013continuous}. If $T$ denotes the absorption time, then
\begin{equation}
 \mathbb P(T>t)\sim t^{-1},\quad p_T(t)\sim t^{-2}.
 \label{eq:duration-bulk}
\end{equation}
We take the avalanche \emph{area} as the primary size observable:
\begin{equation}
 S=\int_0^T n(t) dt.
 \label{eq:area-def}
\end{equation}
Scale invariance then gives $S\sim T^2$ and the joint-density form
\begin{equation}
 p(S,T)=T^{-4}\mathcal F(S/T^2),\quad p_S(S)\sim S^{-3/2}.
 \label{eq:bulk-laws}
\end{equation}

The area is also asymptotically equivalent to the total number of completed events. If $K_-$ denotes the death count at absorption, then its compensator is $S/\tau_0$, and
$
 K_- -\frac{S}{\tau_0}
$
is a stopped counting martingale. On a cutoff-scale excursion with $S=O(N^{4/3})$, its standard deviation is $O(N^{2/3})$, subleading to $S$. The completed-event count and the integrated activity therefore share the same leading edge scaling.

\subsection{Soft-edge balance}

The crossover scale follows directly from Eq.~\eqref{eq:mesoscopic-sde}. Set
$
 n=N^\alpha x,\quad t=N^\alpha s.
$
The branching-noise term requires activity and time to have the same scaling exponent. The cubic drift then carries the factor $N^{3\alpha-2}$, so a nondegenerate balance requires
$
 \alpha=\frac23.
$
The edge variables are therefore
\begin{equation}
 n=N^{2/3}x,
 \quad t=N^{2/3}s,
 \quad \frac{g_N}{\pi\Delta}-1=\kappa N^{-2/3}.
 \label{eq:edge-rescaling}
\end{equation}
The rescaled process converges to Eq.~\eqref{eq:edge-diffusion}. For $\kappa>0$, the deterministic drift vanishes at
$
 x_*=\sqrt{\frac{\kappa}{\beta}},
$
corresponding to the subextensive turning scale $n_*=N^{2/3}x_*$. For $\kappa\leq0$, the same stochastic scaling holds without a positive turning point.

Using Eq.~\eqref{eq:area-def},
\begin{equation}
 S=N^{4/3}\int_0^{\widehat T}X_s ds,
 \quad \widehat T=N^{-2/3}T.
 \label{eq:area-rescaling}
\end{equation}
Hence the finite-size cutoff scales are
\begin{equation}
 T_c\sim N^{2/3},\quad S_c\sim N^{4/3}.
 \label{eq:cutoff-scales}
\end{equation}

The same balance applies when the leading saturation is $-c_p n^p/N^{p-1}$ with $p>1$ and $c_p>0$. In that case,
\begin{equation}
 n_c,T_c\sim N^{(p-1)/p},\quad
 S_c\sim N^{2(p-1)/p}.
 \label{eq:p-class}
\end{equation}
Thus quadratic saturation gives $n_c,T_c\sim N^{1/2}$ and $S_c\sim N$, whereas cubic saturation gives the scales in Eq.~\eqref{eq:cutoff-scales}. The same linear-cubic local balance appears in mean-field tricritical directed percolation \cite{ohtsuki1987nonequilibrium,lubeck2006tricritical}; here we focus on the finite-size cutoff law selected by the Cauchy--Heaviside construction and a fixed microscopic seed.

The contrast with quadratic logistic competition is nevertheless physically significant. For subextensive activity $n\ll N$, the cubic saturation $n^3/N^2$ is smaller than a quadratic term $n^2/N$ by the factor $n/N$. Critical branching excursions are therefore arrested later, at $n_c\sim N^{2/3}$ rather than $N^{1/2}$, and the corresponding avalanche-area cutoff is shifted from the extensive scale $S_c\sim N$ to the scale $S_c\sim N^{4/3}$. The $N^{1/2}$ activity/time edge of a finite-capacity quadratic birth--death process is realized, for example, in the critical susceptible--infected--susceptible (SIS) epidemic \cite{dolgoarshinnykh2006critical},
while logistic Feller diffusions provide the corresponding continuous-state quadratic setting \cite{foucart2025conditioning}. Because the completed-event count is asymptotically equivalent to $S/\tau_0$, the cutoff event count in the repeatable-event lift grows faster than the number of source labels. This is possible because source labels are repeatable; it is not a consequence of Cauchy stability alone. By contrast, an occupancy or refractory constraint, or a response correction, that generates an $n^2/N$ term becomes relevant first and produces the quadratic edge scaling.

\begin{figure*}[t]
 \centering
 \includegraphics[width=\textwidth]{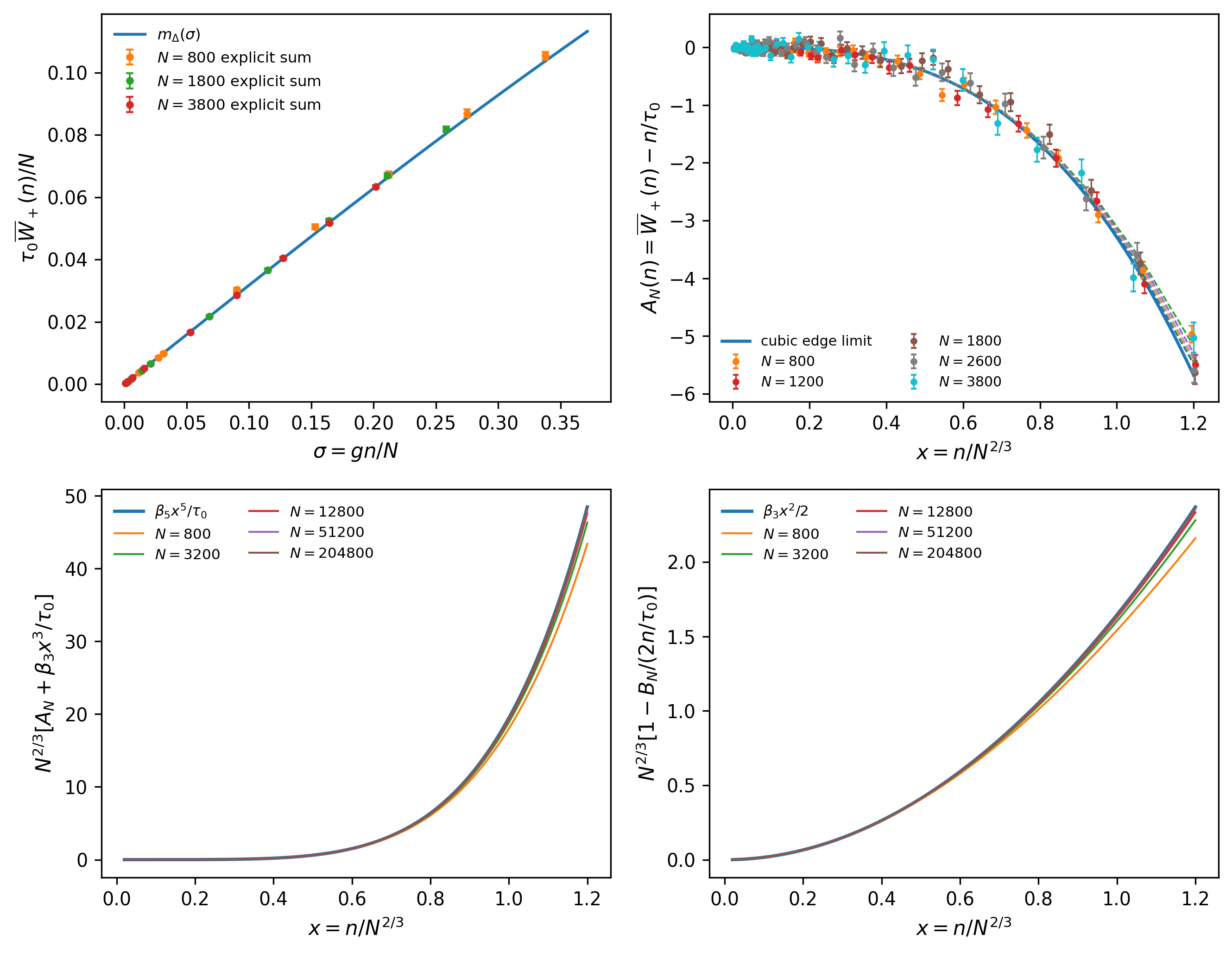}
\caption{\textbf{Exchangeable population generator and convergence to the cubic edge.}
Upper left: threshold response obtained from explicit finite sums of Cauchy event inputs for $N=800,1800,3800$, compared with the exact response $m_\Delta(\sigma)$ in Eq.~\eqref{eq:cauchy-response}. Upper right: population drift expressed in the edge coordinate $x=n/N^{2/3}$ for $N=800,1200,1800,2600,3800$. The finite-$N$ curves use the exact arctangent birth rate Eq.~\eqref{eq:population-birth}, and the limiting curve is the cubic edge drift in Eq.~\eqref{eq:cumulant-class}. Lower left: leading correction to the cubic drift, $N^{2/3}[A_N(n_N)+(\beta/\tau_0)x_N^3]$, with $n_N=\lfloor N^{2/3}x\rfloor$ and $x_N=n_N/N^{2/3}$, approaching $\pi^4x^5/(5\tau_0)$ as predicted by Eq.~\eqref{eq:drift-correction}. Lower right: leading correction to the second jump cumulant, $N^{2/3}[1-B_N(n_N)/(2n_N/\tau_0)]$, approaching $\beta x^2/2$ as in Eq.~\eqref{eq:noise-correction}. The lower panels use the exact rates at $N=800,3200,12800,51200,204800$. The critical comparisons use $\Delta=\tau_0=1$ and $g=\pi$. The explicit-input estimates use $60$ independent  trials per point, and the population-cumulant estimates use $4000$ trials per point; full numerical settings are given in Table~\ref{tab:parameters}.}
 \label{fig:generator-main}
\end{figure*}

\section{Boundary spectral law for the microscopic-seed area measure}
\label{sec:spectral}

The scaling analysis fixes the location of the edge but not its shape. A microscopic avalanche reaches the nonlinear edge only with probability $O(N^{-2/3})$, so the ordinary edge-scale probability mass vanishes as the system grows. We therefore magnify this rare mass rather than normalize it to one. The resulting sigma-finite entrance measure describes how the ordinary branching tail is progressively cut off.

We obtain this measure from the Laplace transform of the area generated by a fixed seed $n_0=O(1)$. Under edge rescaling, the seed approaches the absorbing boundary, so its normalization cannot be inferred from generator convergence started at a finite edge coordinate. Starting from the exact finite-$N$ backward equation, we match the microscopic boundary layer to a universal Feynman--Kac problem and then expose its quartic Dirichlet spectrum by a gauge transformation.

\subsection{Discrete backward equation and edge measure}

Let $\mathbb E_n$ denote expectation for the birth--death process initialized at $n(0)=n$. For a finite-$N$ Laplace parameter $\lambda_N\geq0$, define
$
 M_N(n,\lambda_N)=\mathbb E_n\exp\left(-\lambda_N\int_0^T n(t) dt\right).
$
For the finite-$N$ birth--death process, $M_N$ satisfies the exact backward equation
\begin{align}
 0={}&W^+(n)[M_N(n+1,\lambda_N)-M_N(n,\lambda_N)]\nonumber\\
 &+W^-(n)[M_N(n-1,\lambda_N)-M_N(n,\lambda_N)]\nonumber\\
 &-\lambda_N nM_N(n,\lambda_N),
 \label{eq:discrete-backward}
\end{align}
with $M_N(0,\lambda_N)=1$. For a fixed edge-scale parameter $\widehat\lambda$, set $\lambda_N=\widehat\lambda N^{-4/3}$. Together with $n=N^{2/3}x$, this scaling leads to the Feynman--Kac equation \cite{kac1949distributions} associated with Eq.~\eqref{eq:edge-diffusion}.

For a fixed microscopic seed, the probability mass on the edge scale is $O(N^{-2/3})$. The nontrivial limiting object is therefore the rescaled measure
\begin{equation}
 \mu_N^{(n_0)}(du)=N^{2/3}
 \mathbb P_{n_0}\left(\frac{S}{N^{4/3}}\in du\right).
 \label{eq:edge-measure}
\end{equation}
This measure is not normalized. Its compensated Laplace transform is
\begin{multline}
 \int_0^\infty(1-e^{-\widehat\lambda u})
 \mu_N^{(n_0)}(du)\\
 =N^{2/3}\left[1-M_N(n_0,\widehat\lambda N^{-4/3})\right].
 \label{eq:compensated-transform}
\end{multline}

\subsection{Universal Feynman--Kac problem}

For the edge diffusion, let $T_0$ denote the absorption time, and let $\mathbb E_x$ denote expectation for Eq.~\eqref{eq:edge-diffusion} initialized at $X_0=x$. Define
$
 U_{\widehat\lambda}(x)=\mathbb E_x\exp\left[-\widehat\lambda\int_0^{T_0}X_s ds\right].
$
Its generator is
$
 \mathcal L_\kappa=
 \frac{\kappa x-\beta x^3}{\tau_0}\partial_x+Dx\partial_x^2,
$
and the Feynman--Kac equation is
\begin{equation}
 \mathcal L_\kappa U_{\widehat\lambda}-\widehat\lambda xU_{\widehat\lambda}=0,
 \quad U_{\widehat\lambda}(0)=1.
 \label{eq:FK-dimensional}
\end{equation}
Introduce
\begin{equation}
 \ell=\left(\frac{D\tau_0}{\beta}\right)^{\frac{1}{3}},\quad
 z=\frac{x}{\ell},\quad
 \eta=\frac{\kappa\ell}{D\tau_0},\quad
 \Lambda=\frac{\widehat\lambda\ell^2}{D}.
 \label{eq:metric-factors}
\end{equation}
Writing $U_\Lambda(z)=U_{\widehat\lambda}(\ell z)$ gives
\begin{equation}
 U_\Lambda''+(\eta-z^2)U_\Lambda'-\Lambda U_\Lambda=0,
 \quad U_\Lambda(0)=1,
 \label{eq:universal-FK}
\end{equation}
where boundedness as $z\to\infty$ selects the physical solution.

A fixed seed corresponds to the boundary-scale initial condition
$
 z_0=\frac{n_0}{\ell N^{2/3}}\to0.
$
The associated boundary response is
\begin{equation}
 U_\Lambda(z)=1-z h_\eta(\Lambda)+o(z),
 \quad h_\eta(\Lambda)=-U_\Lambda'(0).
 \label{eq:boundary-response}
\end{equation}
Generator convergence away from $x=0$ is insufficient for this fixed-seed limit. In the Supplemental Material, we stop the discrete process at $m_N=\lfloor\varepsilon N^{2/3}\rfloor$, use the exact birth--death scale function to determine the probability of reaching that level, and show that the Feynman--Kac area accumulated below $m_N$ is negligible after the entrance rescaling. Matching to Eq.~\eqref{eq:boundary-response} then gives
\begin{equation}
 N^{2/3}\left[1-M_N(n_0,\widehat\lambda N^{-4/3})\right]
 \longrightarrow
 \frac{n_0}{\ell}h_\eta\left(\frac{\ell^2}{D}\widehat\lambda\right).
 \label{eq:transform-limit}
\end{equation}
Equation~\eqref{eq:transform-limit} is the central finite-size asymptotic relation. It also fixes the limiting measure, not only its transform. Indeed, for every $\delta>0$,
$
 \mu_N^{(n_0)}([\delta,\infty))
 \leq
 \frac{1}{1-e^{-\widehat\lambda\delta}}
 \int_0^\infty(1-e^{-\widehat\lambda u}) \mu_N^{(n_0)}(du),
$
so Eq.~\eqref{eq:transform-limit} gives uniform local boundedness away from the origin. The standard continuity theorem for Laplace exponents of L\'evy measures then yields vague convergence on $(0,\infty)$,
\begin{equation}
 \mu_N^{(n_0)}\xrightarrow{\mathrm{vague}}
 Q_{\eta,n_0}(u) du,
 \label{eq:vague-convergence}
\end{equation}
where the spectral construction below identifies the density $Q_{\eta,n_0}$. Section~\ref{sec:numerical-main} tests the transform limit directly against exact finite-$N$ backward solutions and tests the same measure at density level.

\subsection{Quartic operator and boundary spectrum}

Set
$
 F_\eta(z)=\eta z-\frac{z^3}{3},\quad
 U_\Lambda(z)=e^{-F_\eta(z)/2}\psi_\Lambda(z).
$
Then
$
 \mathcal H_\eta\psi_\Lambda=-\Lambda\psi_\Lambda,
$
where
\begin{equation}
 \mathcal H_\eta=-\frac{d^2}{dz^2}+\frac14(z^2-\eta)^2-z.
 \label{eq:quartic-operator}
\end{equation}
Let $\mathcal H_\eta^{\rm Dir}$ be the Dirichlet realization of Eq.~\eqref{eq:quartic-operator} on $[0,\infty)$. Its confining potential yields a positive discrete spectrum,
$
 0<E_0(\eta)<E_1(\eta)<\cdots.
$
Let $\chi_j$ be $L^2$-normalized eigenfunctions. Because the microscopic seed collapses onto the absorbing boundary, its mode amplitude is set by boundary flux rather than by a bulk overlap. Define
$
 \rho_j(\eta)=\chi_j'(0)^2.
$
These boundary weights distinguish the entrance law from the spectral expansion generated by a macroscopic initial condition. The defect $w_\Lambda=e^{F_\eta/2}-\psi_\Lambda$, with $\psi_\Lambda(0)=1$, satisfies
$
 w_\Lambda=\Lambda(\mathcal H_\eta^{\rm Dir}+\Lambda)^{-1}e^{F_\eta/2},
 \quad h_\eta(\Lambda)=w_\Lambda'(0).
$
Green's identity gives
$
 \left\langle\chi_j,e^{F_\eta/2}\right\rangle=
 \frac{\chi_j'(0)}{E_j(\eta)},
$
and hence
\begin{equation}
 h_\eta(\Lambda)=\sum_{j=0}^\infty
 \rho_j(\eta)\frac{\Lambda}{E_j(\eta)[E_j(\eta)+\Lambda]}.
 \label{eq:h-spectral}
\end{equation}

Introduce the physical edge variable $u=S/N^{4/3}$ and its dimensionless form
$
 v=\frac{D}{\ell^2}u.
$
The density of the limiting sigma-finite entrance measure in the variable $v$ is
\begin{equation}
 q_\eta(v)=\sum_{j=0}^\infty\rho_j(\eta)e^{-E_j(\eta)v}.
 \label{eq:q-spectral}
\end{equation}
Then
$
 h_\eta(\Lambda)=\int_0^\infty(1-e^{-\Lambda v})q_\eta(v) dv.
$
The singularity $q_\eta(v)\sim v^{-3/2}$ prevents normalization at the origin, while the compensated integral remains finite.

The large-$\Lambda$ boundary layer of Eq.~\eqref{eq:universal-FK} gives $h_\eta(\Lambda)\sim\sqrt{\Lambda}$ and therefore fixes
$
 q_\eta(v)\sim\frac{1}{2\sqrt\pi}v^{-3/2},
 \quad v\downarrow0.
$
Removing this universal branching singularity defines the cutoff function
\begin{equation}
 G_\eta(v)=2\sqrt\pi v^{3/2}q_\eta(v)
 =2\sqrt\pi v^{3/2}\sum_{j=0}^\infty\rho_j(\eta)e^{-E_j(\eta)v}.
 \label{eq:G-spectral}
\end{equation}
By construction, $G_\eta(v)\to1$ as $v\downarrow0$.

Transforming back to the physical edge variable $u$, the limiting density is
\begin{equation}
 Q_{\eta,n_0}(u)=\frac{n_0D}{\ell^3}
 q_\eta\left(\frac{D}{\ell^2}u\right)
 =\frac{n_0}{2\sqrt{\pi D}}u^{-3/2}
 G_\eta\left(\frac{D}{\ell^2}u\right).
 \label{eq:Q-density}
\end{equation}
The accumulation of high-energy modes produces the $v^{-3/2}$ singularity that matches the branching bulk, whereas the lowest modes control the far edge. In particular, the ground state gives the large-$u$ asymptotic
\begin{equation}
 Q_{\eta,n_0}(u)\sim
 \frac{n_0D}{\ell^3}\rho_0(\eta)
 \exp\left[-\frac{E_0(\eta)D}{\ell^2}u\right].
 \label{eq:Q-large-u}
\end{equation}
Thus $E_0$ fixes the far-tail slope, while the full spectrum reconstructs the crossover from the branching bulk to the cutoff.

\begin{figure*}[t]
 \centering
 \includegraphics[width=\textwidth]{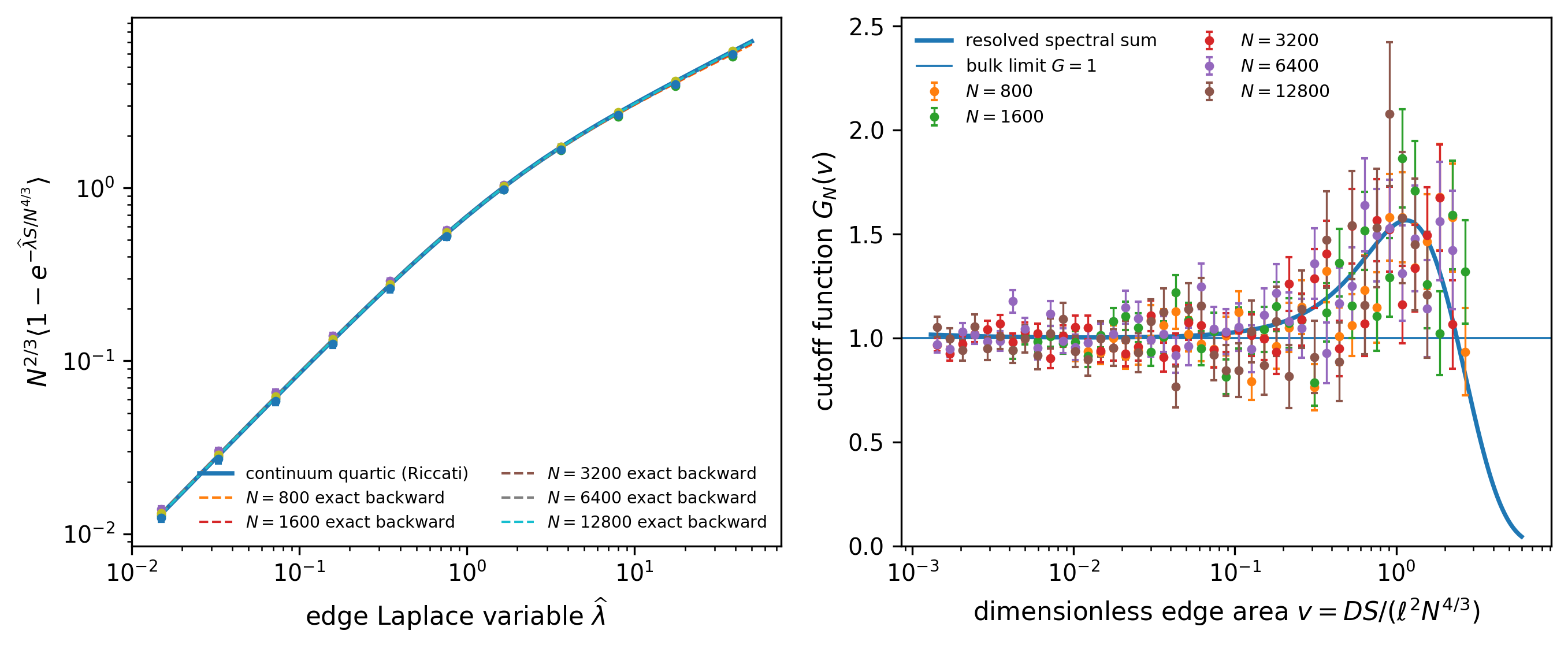}
 \caption{\textbf{Transform- and density-level tests of the microscopic-seed spectral crossover.}
Left: compensated edge transform $N^{2/3}[1-M_N(n_0,\widehat\lambda N^{-4/3})]$ for $N=800,1600,3200,6400,12800$ at the critical point $\eta=0$. The solid curve is the continuum boundary prediction Eq.~\eqref{eq:transform-limit}, evaluated through the quartic Feynman--Kac/Riccati problem; dashed curves are exact finite-$N$ solutions of the discrete backward equation Eq.~\eqref{eq:discrete-backward}, and points are Monte Carlo estimates with standard errors. Right: finite-$N$ cutoff estimator $G_N(v)$ after removal of the universal $v^{-3/2}$ branching singularity. The solid curve is the truncated Dirichlet spectral prediction $G_\eta(v)$ of Eq.~\eqref{eq:G-spectral} over the window resolved by the retained spectrum, and the horizontal line $G=1$ is the branching-bulk limit $v\downarrow0$. All panels use $\Delta=\tau_0=1$ and $n_0=1$; at $\eta=0$ one has $\kappa=0$ and $g_N=\pi$. Trajectory counts, the $\widehat\lambda$ grid, and the spectral truncation parameters are listed in Table~\ref{tab:parameters}.}
 \label{fig:spectral-main}
\end{figure*}
\section{Finite-size tests of the edge theory}
\label{sec:numerical-main}

The discrete process permits several independent tests of the edge theory. Exact moment equations probe the cutoff exponents, the exact rates probe the edge drift and noise, and backward transforms and generalized eigenvalues probe the boundary spectral law. This section summarizes the results; simulation details, fitting windows, and convergence checks are given in the Supplemental Material.

All Monte Carlo simulations use the exact exchangeable Cauchy--Heaviside population rates
\begin{equation}
 W^+(n;g_N)=N m_{\Delta=1}(g_N n/N),\quad W^-(n)=n,
 \label{eq:numerical-rates}
\end{equation}
with $\Delta=\tau_0=1$ and seed $n_0=1$. The critical simulations set $g_N=g_c=\pi$ and span $N=800$ to $12800$. For the critical-window tests at fixed $\eta$, we use
$
 \kappa=\frac{\eta D\tau_0}{\ell},\quad
 g_N=\pi\left(1+\kappa N^{-2/3}\right),
$
as required by Eqs.~\eqref{eq:critical-window-tuning} and \eqref{eq:metric-factors}.

\subsection{Bulk laws and cutoff scales}

Figure~\ref{fig:bulk-edge-main} shows that the same discrete process contains both the branching bulk and the predicted finite-size edge. Fits over successively deeper survival-probability windows approach the BESQ$^0$ slopes $-1/2$ for the area CCDF and $-1$ for the duration CCDF.

Exact moment equations test the cutoff exponents without histogram fitting. Up to $N=6\times10^5$, the local effective exponents of $\langle S^2\rangle/\langle S\rangle$ and $\langle T^3\rangle/\langle T^2\rangle$ converge to $4/3$ and $2/3$, respectively. Figure~\ref{fig:exact-cutoff-exponents} and the Supplemental Material report the finite-size extrapolations and fitting details.

\subsection{Exchangeable generator and finite-size corrections}

The exact arctangent rate permits a direct test of the edge generator, independent of the distributional collapse. Figure~\ref{fig:generator-main} verifies the threshold response, cubic drift, and branching-noise normalization before taking the continuum limit. The lower panels use larger sizes to resolve the leading corrections. Fix $x>0$, let $n_N=\lfloor N^{2/3}x\rfloor$, and define the realized edge coordinate $x_N=n_N/N^{2/3}$, so that $x_N\to x$. Subtracting the cubic term at this realized coordinate gives
\begin{equation}
 N^{2/3}\left[A_N(n_N)+\frac{\beta}{\tau_0}x_N^3\right]
 \longrightarrow \frac{\pi^4}{5\tau_0}x^5.
 \label{eq:drift-correction}
\end{equation}
Likewise,
\begin{equation}
 N^{2/3}\left[1-\frac{B_N(n_N)}{2n_N/\tau_0}\right]
 \longrightarrow \frac{\beta}{2}x^2.
 \label{eq:noise-correction}
\end{equation}
Using $x_N$ in Eq.~\eqref{eq:drift-correction} removes the $O(N^{-2/3})$ lattice-rounding term that would remain if the target coordinate $x$ were substituted before the subtraction. These limits verify the drift and diffusion coefficients entering Eq.~\eqref{eq:edge-diffusion}.

\subsection{Transform and spectral cutoff}

The area transform provides the most direct test of the microscopic-seed boundary law because it can be computed independently by Monte Carlo simulation, by the exact discrete backward equation, and by the continuum Riccati problem. The left panel of Fig.~\ref{fig:spectral-main} compares all three. The logarithmic root-mean-square error (RMSE) between the exact finite-$N$ transform and the continuum limit decreases monotonically with $N$; its fitted exponent, $-0.666$, agrees with the $N^{-2/3}$ correction implied by the quintic term in Eq.~\eqref{eq:Wexpansion}. The right panel removes the universal $v^{-3/2}$ factor and gives a noisier density-level check. Numerical errors and fitting details are reported in the Supplemental Material.

The far-tail spectrum can be tested without rare-event binning. For the finite generator, the area-tail exponents $r_{j,N}$ solve
\begin{equation}
 -\mathcal G_N\varphi=r_{j,N} n\varphi.
 \label{eq:finite-tail-eigenproblem}
\end{equation}
The scaled eigenvalues $\ell^2N^{4/3}r_{j,N}/D$ converge to the quartic eigenvalues $E_j$. At $N=2\times10^5$, the first three modes differ from their continuum values by less than $0.3\%$. Their finite-size dependence is shown in Fig.~\ref{fig:tail-eigenvalues}, with numerical values listed in the Supplemental Material.

Figure~\ref{fig:critical-window} further tests the critical-window family at $\eta=-1,0,1$ for $N=800,3200,12800$. The Supplemental Material lists the spectral values and truncation checks used in these comparisons.

\FloatBarrier

\section{Conclusions}
\label{sec:conclusion}

We have derived a finite-size cutoff theory for an exchangeable repeatable-event process whose hydrodynamic drift is fixed by the low-activity Cauchy DMFT. The DMFT determines the linear-cubic drift, whereas the event process supplies absorbing birth--death fluctuations. At criticality, small avalanches lie in the squared-Bessel branching regime. Cubic saturation becomes relevant only at activity and duration scales of order $N^{2/3}$, producing an avalanche-area cutoff of order $N^{4/3}$. Relative to quadratic logistic saturation, the cubic term is weaker throughout the subextensive regime and therefore delays finite-size arrest.

For a fixed microscopic seed, the edge limit is a sigma-finite entrance measure rather than a normalized probability distribution. A discrete scale-function matching calculation fixes its boundary normalization, and the compensated-transform limit determines the measure vaguely away from the absorbing boundary. The transform is the boundary response of a quartic Schr\"odinger operator on the half-line. The Dirichlet eigenvalues determine the decay scales, and the squared boundary derivatives of the eigenfunctions determine their weights. This boundary-flux weighting, rather than a bulk spectral overlap, is the central finite-size distinction of the microscopic-seed problem. The resulting spectral representation contains both limits of the distribution: the accumulation of high modes reproduces the branching singularity, while the ground state determines the exponential far tail. Exact moments, backward transforms, generalized eigenvalues, and Monte Carlo simulations consistently support this structure.

The exchangeable population process is a mesoscopic realization of the Cauchy mean-field mode, not a pathwise reduction of a fixed quenched network. In a quenched system, dynamically selected source labels may correlate with atypical coupling columns, so the event count need not remain Markovian. The finite-size class also depends on the microscopic lift. Any occupancy, refractory, or response mechanism that generates a quadratic correction at low activity becomes relevant before the cubic edge. The spectral law therefore applies to processes satisfying both the edge-cumulant convergence and the near-boundary rate conditions stated in the Supplemental Material.

The experimentally distinguishable prediction is not a bulk exponent alone, but the joint occurrence of mean-field branching statistics, $N^{2/3}$ duration scaling, $N^{4/3}$ area scaling, and the dimensionless spectral cutoff function $G_\eta$. In data, the effective system size need not equal the number of recorded units: finite observation windows, spatial subsampling, measurement overlap, and temporal binning can all modify the observed cutoff \cite{Yu2014,priesemann2009subsampling,levina2017subsampling,neto2022sampling,pausch2022neuronal}.

\bibliographystyle{apsrev4-2}
\bibliography{avalanche}

\clearpage
\onecolumngrid
\setcounter{equation}{0}
\setcounter{figure}{0}
\setcounter{table}{0}
\setcounter{section}{0}
\renewcommand{\theequation}{S\arabic{equation}}
\renewcommand{\thefigure}{S\arabic{figure}}
\renewcommand{\thetable}{S\arabic{table}}
\renewcommand{\thesection}{\Roman{section}}
\renewcommand{\thesubsection}{\Alph{subsection}}
\renewcommand{\theHequation}{S\arabic{equation}}
\renewcommand{\theHfigure}{S\arabic{figure}}
\renewcommand{\theHtable}{S\arabic{table}}
\renewcommand{\theHsection}{S\Roman{section}}

\begin{center}
\textbf{\large Supplemental Material for:}\par
\vspace{0.35em}
\textbf{\large Finite-size scaling and cutoff functions of critical avalanches under cubic saturation}\par
\vspace{0.55em}
Ryota Kojima, Kazufumi Hosoda, Hiroyuki Suto, and Hitoshi Yamada
\end{center}
\vspace{0.5em}

\section{Exchangeable generator and finite-size class}
\label{app:generator}

For a fixed cardinality $n$, Eq.~\eqref{eq:averaged-rate-def} gives
\begin{equation}
 W^+(n)=\frac{N}{\tau_0}m_\Delta(gn/N),\quad
 W^-(n)=\frac{n}{\tau_0},
 \label{eq:app-rates}
\end{equation}
with
\begin{equation}
 m_\Delta(\sigma)=\frac{1}{\pi}\arctan\frac{\sigma}{\Delta}.
\end{equation}
Let $P(n,t)=\mathbb P[n(t)=n]$. The master equation is
\begin{align}
 \partial_tP(n,t)=W^+(n-1)P(n-1,t)+W^-(n+1)P(n+1,t)
 -[W^+(n)+W^-(n)]P(n,t).
 \label{eq:app-master}
\end{align}
Expanding at low activity gives Eq.~\eqref{eq:Wexpansion}; at criticality,
\begin{equation}
 A_N(n)=-\frac{\pi^2}{3\tau_0}\frac{n^3}{N^2}
 +\frac{\pi^4}{5\tau_0}\frac{n^5}{N^4}+\cdots,
 \quad B_N(n)=\frac{2n}{\tau_0}+o(n).
 \label{eq:app-cumulants}
\end{equation}

For an exclusion process,
\begin{equation}
 W^+_{\mathrm{occ}}(n)=\frac{N-n}{\tau_0}m_\Delta(gn/N),
\end{equation}
and
\begin{equation}
 W^+_{\mathrm{occ}}(n)-W^-(n)=\frac1{\tau_0}\left[
 \left(\frac{g}{\pi\Delta}-1\right)n
 -\frac{g}{\pi\Delta}\frac{n^2}{N}
 -\frac{g^3}{3\pi\Delta^3}\frac{n^3}{N^2}+\cdots\right].
\end{equation}
The $n^2/N$ correction becomes relevant before the cubic term and therefore defines a different finite-size class.

\section{Generator convergence and microscopic-seed matching}
\label{app:generator-convergence}

\subsection{Edge limit and boundary assumptions}

Let $f_N(n)=\varphi(nN^{-2/3})$ and $x=nN^{-2/3}$. Taylor expansion of Eq.~\eqref{eq:discrete-generator} gives
\begin{align}
 \mathcal G_Nf_N(n)
 =& A_N(n)N^{-2/3}\varphi'(x)
 +\frac12B_N(n)N^{-4/3}\varphi''(x)\nonumber\\
 &+\frac16A_N(n)N^{-2}\varphi'''(x)
 +O \left([W^+(n)+W^-(n)]N^{-8/3}\right).
 \label{eq:app-generator-expansion}
\end{align}
Under the tuning in Eq.~\eqref{eq:critical-window-tuning}, multiplication by $N^{2/3}$ yields Eq.~\eqref{eq:generator-convergence}. Every term of third order or higher is $O(N^{-2/3})$ on a fixed interval $x\in[\varepsilon,L]\subset(0,\infty)$.

For a general single-step process, convergence from an edge-scale initial state requires
\begin{align}
 \sup_{x\in[\varepsilon,L]}
 \left|A_N(\lfloor N^{2/3}x\rfloor)
 -\frac{\kappa x-\beta x^3}{\tau_0}\right|&\longrightarrow0,
 \label{eq:app-edge-A}\\
 \sup_{x\in[\varepsilon,L]}
 \left|N^{-2/3}B_N(\lfloor N^{2/3}x\rfloor)-2Dx\right|&\longrightarrow0,
 \label{eq:app-edge-B}
\end{align}
for every $0<\varepsilon<L<\infty$, together with vanishing rescaled higher jump cumulants. For
\begin{equation}
 \lambda_N=\widehat\lambda N^{-4/3},
 \label{eq:app-lambdaN}
\end{equation}
the generator convergence above, applied to chains stopped on compact edge intervals, gives for each fixed $x>0$
\begin{equation}
 M_N(\lfloor N^{2/3}x\rfloor,\lambda_N)
 \longrightarrow U_{\widehat\lambda}(x)
 =U_\Lambda(x/\ell),
 \quad \Lambda=\frac{\ell^2}{D}\widehat\lambda.
 \label{eq:app-outer-transform-limit}
\end{equation}
The negative cubic drift permits removal of the upper stopping level as $L\to\infty$.

Equation~\eqref{eq:app-outer-transform-limit} does not determine the normalization generated by a fixed seed. To control the boundary layer, assume that for some $x_b>0$,
\begin{align}
 \sup_{1\leq n\leq x_bN^{2/3}}
 \left|\frac{B_N(n)}{2Dn}-1\right|&\longrightarrow0,
 \label{eq:app-boundary-B}\\
 \sup_{1\leq n\leq x_bN^{2/3}}
 \left|
 N^{2/3}\log\frac{W^-(n)}{W^+(n)}
 +\frac{\kappa-\beta(n/N^{2/3})^2}{D\tau_0}
 \right|&\longrightarrow0.
 \label{eq:app-boundary-ratio}
\end{align}
For the Cauchy--Heaviside rates, these conditions follow from the uniform expansion, valid for $0<n\leq x_bN^{2/3}$,
\begin{align}
 W^+(n)
 &=\frac{n}{\tau_0}
 \left[1+N^{-2/3}(\kappa-\beta x^2)+O(N^{-4/3})\right],\nonumber\\
 W^-(n)&=\frac{n}{\tau_0},
 \quad x=\frac{n}{N^{2/3}},
 \label{eq:app-boundary-expansion}
\end{align}
with $D=1/\tau_0$.

\subsection{Matching from a fixed seed}

Fix $0<\varepsilon<x_b$ and set
\begin{equation}
 m_N=\lfloor\varepsilon N^{2/3}\rfloor.
\end{equation}
Let $\vartheta_a=\inf\{t\geq0:n(t)=a\}$ be the first hitting time of state $a$, and define
\begin{equation}
 p_N(n,m)=\mathbb P_n(\vartheta_m<\vartheta_0).
\end{equation}
The exact birth--death scale formula is
\begin{equation}
 p_N(n,m)=
 \frac{\displaystyle\sum_{k=0}^{n-1}R_{N,k}}
 {\displaystyle\sum_{k=0}^{m-1}R_{N,k}},
 \quad
 R_{N,0}=1,\quad
 R_{N,k}=\prod_{j=1}^{k}\frac{W^-(j)}{W^+(j)}.
 \label{eq:app-hitting-probability}
\end{equation}
For $k=\lfloor xN^{2/3}\rfloor$, Eq.~\eqref{eq:app-boundary-ratio} gives
\begin{equation}
 \log R_{N,k}\longrightarrow
 -\frac{\kappa x}{D\tau_0}
 +\frac{\beta x^3}{3D\tau_0}.
 \label{eq:app-scale-product-limit}
\end{equation}
Hence, with
\begin{equation}
 \mathfrak s_\kappa(\varepsilon)=
 \int_0^\varepsilon
 \exp \left[-\frac{\kappa y}{D\tau_0}
 +\frac{\beta y^3}{3D\tau_0}\right]dy,
 \label{eq:app-scale-function}
\end{equation}
Riemann-sum convergence in the denominator, together with $R_{N,k}\to1$ for each fixed $k$ in the numerator, yields
\begin{equation}
 N^{2/3}p_N(n_0,m_N)
 \longrightarrow
 \frac{n_0}{\mathfrak s_\kappa(\varepsilon)}.
 \label{eq:app-hitting-limit}
\end{equation}

We next bound the area accumulated before the process reaches either $0$ or $m_N$. Define
\begin{equation}
 \mathcal A_m=
 \int_0^{\vartheta_0\wedge\vartheta_m}n(t) dt.
\end{equation}
For $V_m(n)=n(m-n)$,
\begin{equation}
 -\mathcal G_NV_m(n)
 =B_N(n)-A_N(n)(m-2n).
 \label{eq:app-barrier-generator}
\end{equation}
For sufficiently small $\varepsilon$, Eqs.~\eqref{eq:app-boundary-B} and \eqref{eq:app-boundary-ratio} imply the existence of $c>0$, independent of $N$, such that
\begin{equation}
 -\mathcal G_NV_m(n)\geq cn,
 \quad 1\leq n<m\leq\varepsilon N^{2/3}.
\end{equation}
Dynkin's formula then gives
\begin{equation}
 \mathbb E_n\mathcal A_m
 \leq\frac{n m}{c}.
 \label{eq:app-inner-area-bound}
\end{equation}

The strong Markov property at $\vartheta_0\wedge\vartheta_{m_N}$, together with $0\leq1-e^{-y}\leq y$, gives
\begin{equation}
 0\leq
 [1-M_N(n_0,\lambda_N)]
 -p_N(n_0,m_N)[1-M_N(m_N,\lambda_N)]
 \leq \lambda_N\mathbb E_{n_0}\mathcal A_{m_N}.
 \label{eq:app-matching-inequality}
\end{equation}
Multiplying by $N^{2/3}$ and using Eq.~\eqref{eq:app-inner-area-bound} bounds the right-hand side by
\begin{equation}
 \frac{\widehat\lambda n_0}{c}\varepsilon+o(1).
\end{equation}
At fixed $\varepsilon$, Eqs.~\eqref{eq:app-outer-transform-limit} and \eqref{eq:app-hitting-limit} therefore give
\begin{equation}
 \limsup_{N\to\infty}
 \left|
 N^{2/3}[1-M_N(n_0,\lambda_N)]
 -\frac{n_0}{\mathfrak s_\kappa(\varepsilon)}
 [1-U_\Lambda(\varepsilon/\ell)]
 \right|
 \leq C\widehat\lambda n_0\varepsilon,
 \label{eq:app-matched-limit}
\end{equation}
where $C$ can be chosen uniformly for sufficiently small $\varepsilon$. Finally,
\begin{equation}
 \mathfrak s_\kappa(\varepsilon)=\varepsilon+O(\varepsilon^2),
 \quad
 1-U_\Lambda(\varepsilon/\ell)
 =\frac{\varepsilon}{\ell}h_\eta(\Lambda)+o(\varepsilon).
\end{equation}
Taking first $N\to\infty$ and then $\varepsilon\downarrow0$ proves
\begin{equation}
 N^{2/3}
 [1-M_N(n_0,\widehat\lambda N^{-4/3})]
 \longrightarrow
 \frac{n_0}{\ell}
 h_\eta \left(\frac{\ell^2}{D}\widehat\lambda\right),
 \label{eq:app-microscopic-seed-limit}
\end{equation}
which establishes Eq.~\eqref{eq:transform-limit}.

For completeness, the transform limit also implies convergence of the rescaled edge measures. For any fixed $\delta>0$ and $\widehat\lambda>0$,
\begin{equation}
 \mu_N^{(n_0)}([\delta,\infty))
 \leq
 \frac{N^{2/3}[1-M_N(n_0,\widehat\lambda N^{-4/3})]}
 {1-e^{-\widehat\lambda\delta}},
 \label{eq:app-local-measure-bound}
\end{equation}
so Eq.~\eqref{eq:app-microscopic-seed-limit} gives uniform local boundedness on $(0,\infty)$. Moreover, finiteness of the compensated transform is equivalent, up to constants, to
$\int(1\wedge u) \mu_N^{(n_0)}(du)<\infty$.
The continuity theorem for Laplace exponents of L\'evy measures therefore identifies every vague subsequential limit from its compensated transform. Since Eq.~\eqref{eq:app-microscopic-seed-limit} holds for every $\widehat\lambda>0$, the limit is unique and equals the measure with density in Eq.~\eqref{eq:Q-density}. This proves Eq.~\eqref{eq:vague-convergence}.

\section{Bulk scaling, exact moments, and event counts}
\label{app:bulk-moments}

For the critical Feller diffusion, scale invariance gives
\begin{equation}
 p(S,T)=T^{-4}\mathcal F(S/T^2),
\end{equation}
and integration gives Eq.~\eqref{eq:bulk-laws}. For the $N=12800$ bulk-law fits, the avalanche realizations were partitioned into 160 contiguous blocks in simulation order. We generated 800 bootstrap replicates by resampling these blocks with replacement. The fits were performed over ranges in which the empirical complementary cumulative distribution function decreased from $0.2$ to $0.02$, from $0.15$ to $0.01$, and from $0.1$ to $0.005$. These fits give the area-CCDF slopes
\begin{equation}
 -0.5025,\quad -0.5027,\quad -0.5066,
\end{equation}
with respective $95\%$ confidence intervals $[-0.5115,-0.4944]$, $[-0.5121,-0.4934]$, and $[-0.5190,-0.4948]$. The corresponding duration slopes are
\begin{equation}
 -0.9361,\quad -0.9577,\quad -0.9802,
\end{equation}
with intervals $[-0.9503,-0.9228]$, $[-0.9763,-0.9388]$, and $[-1.0031,-0.9563]$. The preasymptotic bias in the duration slope decreases as the fitting window moves farther into the tail.

For the discrete process, define
\begin{equation}
 m_k(n)=\mathbb E_n[S^k],\quad t_k(n)=\mathbb E_n[T^k].
\end{equation}
With $m_0=t_0=1$, the exact recursions are
\begin{equation}
 -\mathcal G_Nm_k(n)=kn m_{k-1}(n),\quad m_k(0)=0,
 \label{eq:area-moments}
\end{equation}
\begin{equation}
 -\mathcal G_Nt_k(n)=k t_{k-1}(n),\quad t_k(0)=0.
 \label{eq:duration-moments}
\end{equation}
We solved these tridiagonal systems for 19 sizes between $N=400$ and $N=600000$. The moment ratios isolate the upper cutoff:
\begin{equation}
 \frac{\langle S^2\rangle}{\langle S\rangle}\sim N^{4/3},
 \quad
 \frac{\langle T^3\rangle}{\langle T^2\rangle}\sim N^{2/3}.
\end{equation}
We computed local exponents from adjacent logarithmic slopes and assigned them to $N_{{\rm mid},i}=\sqrt{N_iN_{i+1}}$. At the largest sizes, the local exponents reach $1.33293$ and $0.66613$. Linear fits in $N_{\rm mid}^{-2/3}$ over the eight largest-$N$ intervals extrapolate to $1.33333$ and $0.66650$.

\begin{figure}[!htbp]
 \centering
 \includegraphics[width=0.82\textwidth]{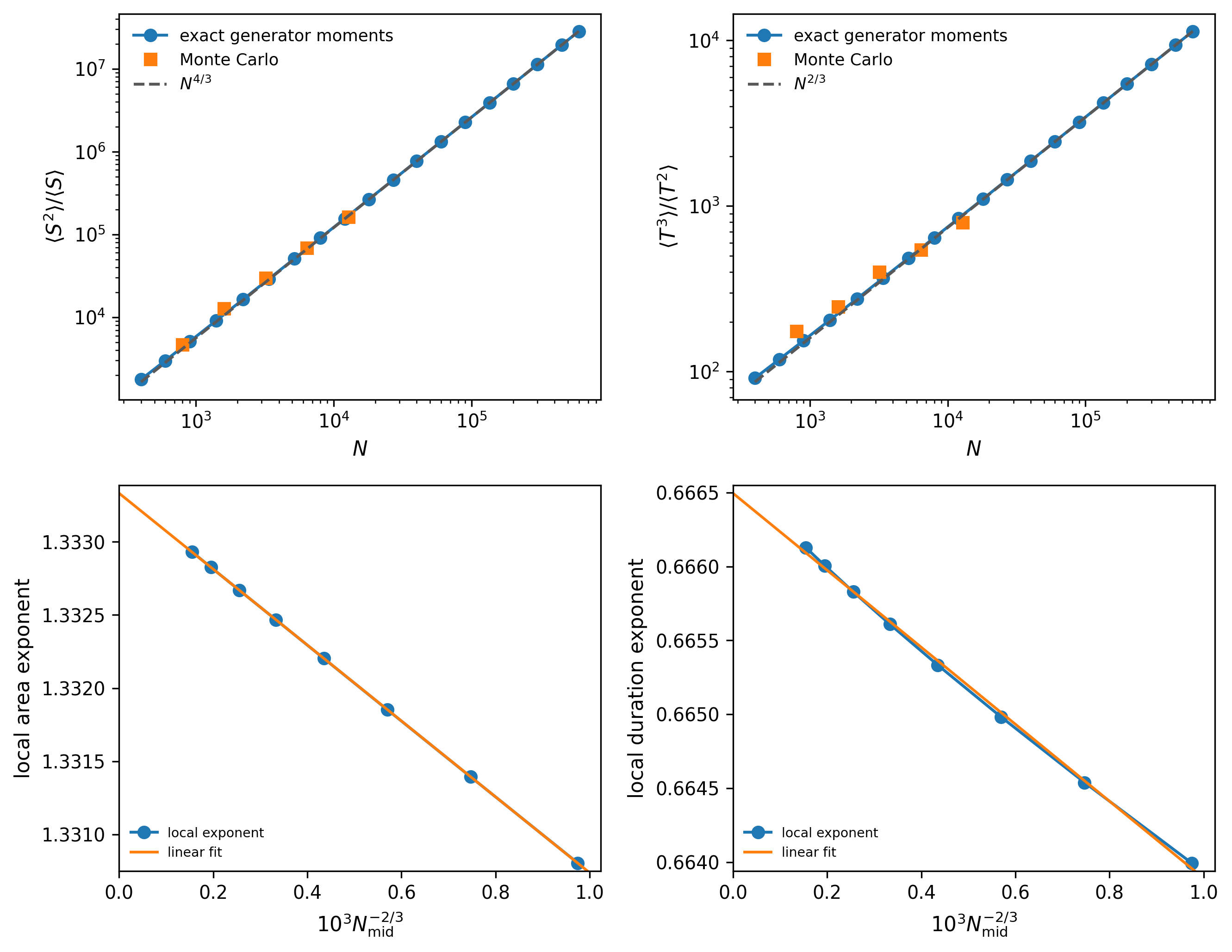}
\caption{\textbf{Exact finite-size cutoff exponents from moment equations.}
Upper panels: cutoff-sensitive moment ratios $\langle S^2\rangle/\langle S\rangle$ (left) and $\langle T^3\rangle/\langle T^2\rangle$ (right), obtained from the exact finite-$N$ recursions in Eqs.~\eqref{eq:area-moments} and \eqref{eq:duration-moments}. The reference laws are $N^{4/3}$ for the area ratio and $N^{2/3}$ for the duration ratio; Monte Carlo values provide independent checks. Exact calculations use $19$ system sizes between $N=400$ and $N=600000$. Lower panels: local effective exponents from adjacent logarithmic slopes, plotted against $10^3N_{\rm mid}^{-2/3}$ with $N_{{\rm mid},i}=\sqrt{N_iN_{i+1}}$ for the eight largest-$N$ intervals. Linear extrapolation to $N^{-2/3}\to0$ gives intercepts $1.33333$ for the area cutoff and $0.66650$ for the duration cutoff, consistent with $4/3$ and $2/3$, respectively. The calculations are at criticality with $\Delta=\tau_0=1$, $n_0=1$, and $g=\pi$.}
 \label{fig:exact-cutoff-exponents}
\end{figure}

Let $K_-(t)$ denote the death counting process. Its stochastic intensity is $n(t)/\tau_0$, so
\begin{equation}
 K_-(T)-\frac{S}{\tau_0}
\end{equation}
is a martingale stopped at absorption. On the event scale $S=O(N^{4/3})$, its standard deviation is $O(N^{2/3})$, so the completed-event count and $S/\tau_0$ have the same leading edge limit.

\section{Boundary spectral representation}
\label{app:spectral}

The gauge transform factorizes the Dirichlet operator as
\begin{equation}
 \mathcal H_\eta=
 \left(\partial_z+\frac{\eta-z^2}{2}\right)
 \left(-\partial_z+\frac{\eta-z^2}{2}\right)
 =
 -\partial_z^2+\frac14(z^2-\eta)^2-z.
\end{equation}
The factorization makes the spectrum nonnegative. The zero mode fails the Dirichlet boundary condition, and hence $E_j>0$.

For $L^2$-normalized Dirichlet eigenfunctions $\chi_j$, Green's identity gives
\begin{align}
 E_j\left\langle\chi_j,e^{F_\eta/2}\right\rangle
 &=\chi_j'(0),\\
 w_\Lambda
 &=\Lambda\sum_j
 \frac{\left\langle\chi_j,e^{F_\eta/2}\right\rangle}
 {E_j+\Lambda}\chi_j.
\end{align}
Differentiating at the boundary yields Eq.~\eqref{eq:h-spectral}.

For each mode,
\begin{equation}
 \int_0^\infty(1-e^{-\Lambda v})e^{-E_jv} dv
 =\frac{\Lambda}{E_j(E_j+\Lambda)},
\end{equation}
which establishes Eqs.~\eqref{eq:q-spectral}--\eqref{eq:G-spectral}. Equation~\eqref{eq:Q-density} follows by setting $v=Du/\ell^2$ and using the boundary expansion at $z_0=n_0/(\ell N^{2/3})$.

The large-area decay rates can also be extracted directly at finite $N$. Using the area clock $a(t)=\int_0^t n(s) ds$, the time-changed generator is $n^{-1}\mathcal G_N$ for $n>0$. Its killed eigenvalue equation is therefore the generalized problem in Eq.~\eqref{eq:finite-tail-eigenproblem}, and edge scaling predicts
\begin{equation}
 \frac{\ell^2}{D}N^{4/3}r_{j,N}\longrightarrow E_j.
 \label{eq:tail-convergence}
\end{equation}

\begin{figure}[!htbp]
 \centering
 \includegraphics[width=0.76\textwidth]{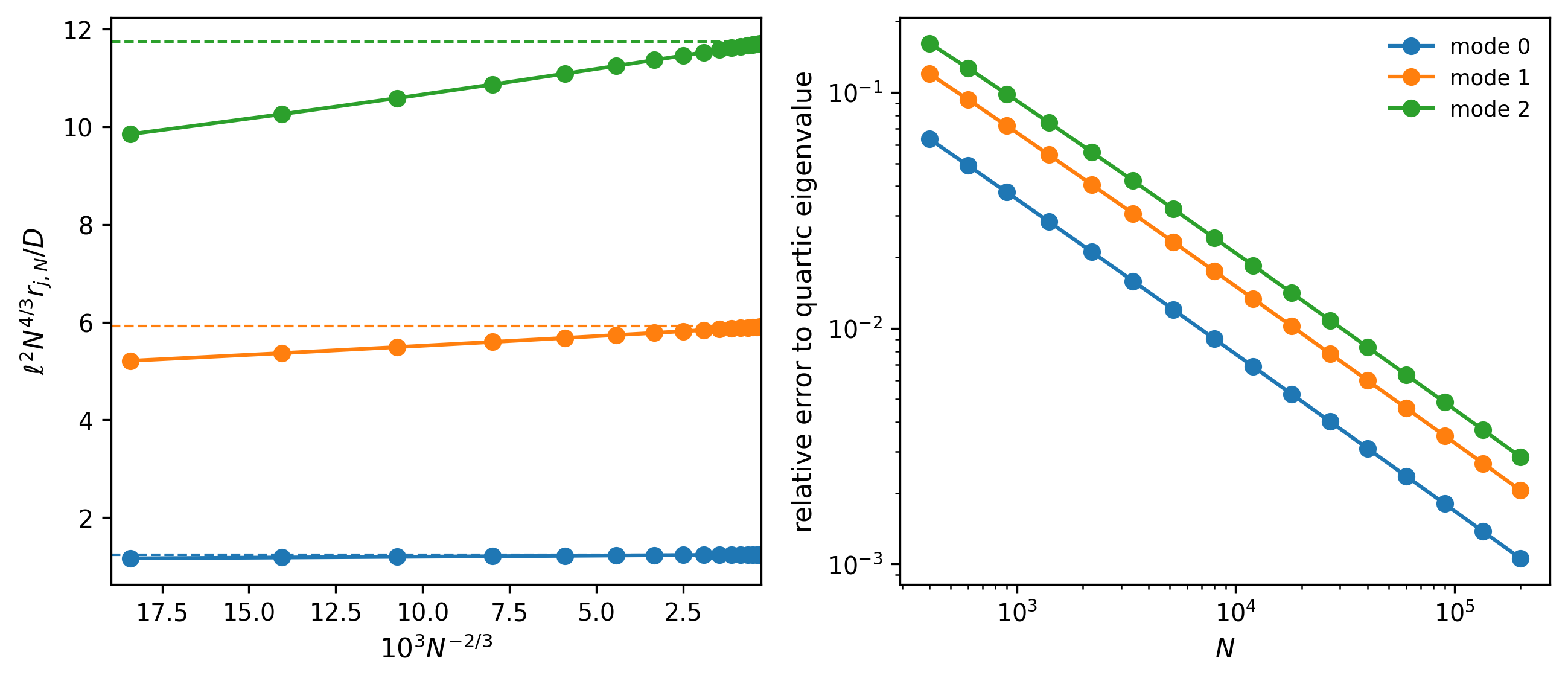}
\caption{\textbf{Direct finite-$N$ convergence of the area-tail spectrum.}
Left: the first three generalized finite-$N$ area-tail eigenvalues $r_{j,N}$ from Eq.~\eqref{eq:finite-tail-eigenproblem}, shown in the edge-scaled form $(\ell^2/D)N^{4/3}r_{j,N}$ versus $10^3N^{-2/3}$. The calculation uses $16$ system sizes from $N=400$ to $N=200000$. Horizontal dashed lines are the corresponding continuum quartic Dirichlet eigenvalues $E_j$, to which the scaled finite-$N$ values converge according to Eq.~\eqref{eq:tail-convergence}. Right: relative errors of the same three modes with respect to their continuum values. At $N=200000$, the relative errors of the first three modes are $1.06\times10^{-3}$, $2.06\times10^{-3}$, and $2.85\times10^{-3}$. The comparison is for the critical theory $\eta=0$ with $\Delta=\tau_0=1$.}
 \label{fig:tail-eigenvalues}
\end{figure}

\section{Numerical methods}
\label{app:numerics}

\subsection{Simulation and generator validation}

We generated continuous-time trajectories with the Gillespie algorithm using Eq.~\eqref{eq:numerical-rates}. At state $n>0$, the holding time is exponential with rate $W^+(n;g_N)+W^-(n)$, and the next jump is a birth with probability $W^+(n;g_N)/[W^+(n;g_N)+W^-(n)]$. The area was accumulated exactly over each holding interval. A hard guard of $5\times10^7$ jumps per avalanche was imposed; reaching this guard was treated as a numerical failure and aborted the run rather than being recorded as an extinction event. None of the reported simulations triggered this guard.

Because the fixed-seed edge mass decreases as $N^{-2/3}$, the number of critical trajectories was increased approximately as $N^{2/3}$, starting from $40000$ at $N=800$, with a computational cap of $120000$ trajectories per system size. This gives $40000$, $63000$, $101000$, $120000$, and $120000$ avalanches for $N=800,1600,3200,6400,12800$, respectively. These critical runs use $g_N=\pi$. The detuned transform tests used $N=800,3200,12800$ with $8000$, $20000$, and $35000$ avalanches for each of $\eta=-1,0,1$; their trajectory counts were likewise increased approximately as $N^{2/3}$ from the $N=800$ value and capped at $35000$. For each detuned run, $\kappa=\eta D\tau_0/\ell$ and $g_N=\pi(1+\kappa N^{-2/3})$.

For the explicit-input test, we used the distributionally equivalent independent-summand representation
\begin{equation}
 I_i=h+\sum_{r=1}^nJ_{ir},\quad J_{ir}\sim\mathrm{Cauchy}(0,g/N).
\end{equation}
At fixed $n$, Cauchy stability gives the same one-target input law as the repeated-label sum in Eq.~\eqref{eq:event-input}, because repeated labels enter there with positive integer multiplicities whose total is $n$. We thresholded these samples for $N=800,1800,3800$, using six edge-coordinate values together with small-$n$ checks and 60 independent trials per point. The population-cumulant estimates used direct Cauchy sampling with 4000 trials per point at $N=800,1200,1800,2600,3800$. The deterministic correction panels in Fig.~\ref{fig:generator-main} used the exact population rates at $N=800,3200,12800,51200,204800$.

\begin{table}[!htbp]
\caption{Numerical parameters.}
\label{tab:parameters}
\begin{ruledtabular}
\begin{tabular}{ll}
Quantity & Value\\
\hline
Critical sizes & $800,1600,3200,6400,12800$\\
Critical avalanches & $40000,63000,101000,120000,120000$\\
Detuned sizes & $800,3200,12800$\\
Detuned avalanches & $8000,20000,35000$ per $\eta$\\
Detunings & $\eta=-1,0,1$\\
Exact-moment sizes & $400$--$600000$ (19 sizes)\\
Tail-eigenvalue sizes & $400$--$200000$ (16 sizes)\\
Explicit-sum sizes & $800,1800,3800$\\
Explicit trials per point & $60$\\
Population-cumulant sizes & $800,1200,1800,2600,3800$\\
Population-cumulant trials & $4000$ per point\\
Asymptotic-correction sizes & $800,3200,12800,51200,204800$\\
Transform grid & $32$ points, $0.015\leq\widehat\lambda\leq50$\\
Bootstrap & $800$ replicates, $160$ contiguous blocks\\
Jump guard & $5\times10^7$ jumps per avalanche\\
Spectral domain & $z_{\max}=10$\\
Spectral grid intervals & $N_z=2400$\\
Retained modes & $N_{\rm modes}=240$\\
Minimum histogram count & $15$\\
Entrance threshold & $S_{\min}=30$\\
Resolved-window constant & $c_{\mathrm{res}}=8$\\
Random seed & $20260520$\\
\end{tabular}
\end{ruledtabular}
\end{table}

\subsection{Transform calculations}

We solved the tridiagonal discrete backward equation with upper truncation $n_{\max}=3N$. Repeating the calculation at $2N$, $3N$, and $4N$ changed the reported transform by less than numerical precision. For the continuum transform, we introduced the logarithmic derivative $r(z)=\psi_\Lambda'(z)/\psi_\Lambda(z)$, which obeys
\begin{equation}
 r'(z)=\frac14(z^2-\eta)^2-z+\Lambda-r(z)^2,
 \quad h_\eta(\Lambda)=\frac{\eta}{2}-r(0).
\end{equation}
The solution compatible with bounded $U_\Lambda$ has the large-$z$ asymptotic
\begin{equation}
 r(z)=\frac{\eta-z^2}{2}-\frac{\Lambda}{z^2}+O(z^{-4}),
\end{equation}
which supplies the outer boundary condition for numerical integration from the large-$z$ end of the interval. The logarithmic RMSE is the root mean square of the pointwise differences between the logarithms of the discrete and continuum compensated transforms over the 32-point $\widehat\lambda$ grid. Its values for $N=800,1600,3200,6400,12800$ are $0.01573$, $0.00992$, $0.00625$, $0.00394$, and $0.00248$, respectively. A log--log fit over these five sizes gives an exponent of $-0.666$, consistent with the predicted $N^{-2/3}$ correction.

\begin{figure}[!htbp]
 \centering
 \includegraphics[width=0.70\textwidth]{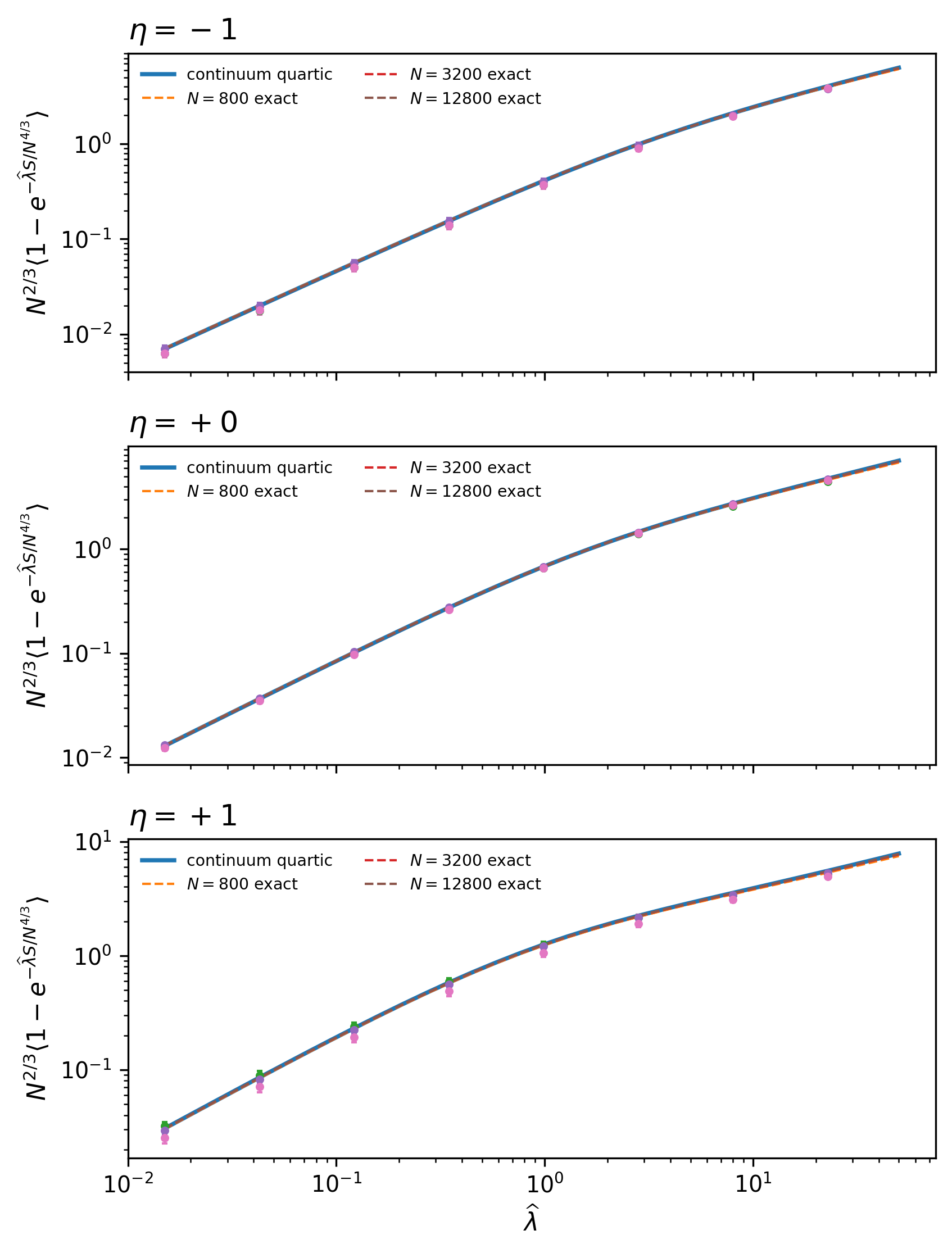}
\caption{\textbf{Critical-window crossover away from exact criticality.}
Compensated microscopic-seed area transforms for $\eta=-1,0,1$ at $N=800,3200,12800$. For each detuning, $\kappa=\eta D\tau_0/\ell$ and $g_N=\pi[1+\kappa N^{-2/3}]$, as required by Eqs.~\eqref{eq:critical-window-tuning} and \eqref{eq:metric-factors}. Solid curves are the continuum quartic boundary predictions from Eq.~\eqref{eq:transform-limit}; dashed curves are exact finite-$N$ solutions of the backward equation Eq.~\eqref{eq:discrete-backward}; points are Monte Carlo estimates. All calculations use $\Delta=\tau_0=1$ and the fixed microscopic seed $n_0=1$. The Monte Carlo sample counts are $8000$, $20000$, and $35000$ avalanches for $N=800$, $3200$, and $12800$, respectively, for each value of $\eta$.}
 \label{fig:critical-window}
\end{figure}

\subsection{Spectral discretization and finite-size tail}

We discretized the operator on $[0,z_{\max}]$ using a centered second derivative and Dirichlet endpoints. The spectral calculation used
\begin{equation}
 z_{\max}=10,\quad N_z=2400,\quad N_{\rm modes}=240
\end{equation}
retained eigenpairs. After continuum normalization of the eigenvectors,
\begin{equation}
 \rho_j=[\chi_j'(0)]^2,
\end{equation}
where the boundary derivative was evaluated with a second-order one-sided formula. The resulting values are
\begin{align}
 E_0(-1)&=2.2438878,& \rho_0(-1)&=2.0575254,\\
 E_0(0)&=1.2407073,& \rho_0(0)&=1.4731256,\\
 E_0(1)&=0.5906898,& \rho_0(1)&=0.9363436.
\end{align}
We solved the generalized finite-$N$ tail problem for 16 sizes up to $N=200000$. At the largest size, the first three scaled eigenvalues are $1.2393949$, $5.9058102$, and $11.712617$, compared with the continuum values $1.2407073$, $5.9179766$, and $11.746090$. The corresponding relative errors are $1.06\times10^{-3}$, $2.06\times10^{-3}$, and $2.85\times10^{-3}$. Varying the finite-state truncation from $2N$ to $4N$ changes the scaled eigenvalues by at most $2.3\times10^{-11}$ and the exact moments by less than $1.2\times10^{-9}$.

\subsection{Cutoff-density estimator}

For each trajectory, the dimensionless area introduced in Sec.~\ref{sec:spectral} is
\begin{equation}
 v=\frac{DS}{\ell^2N^{4/3}}.
\end{equation}
Let $p_{N,v}(v)$ be the probability density of the dimensionless area for seed $n_0$. We used the same logarithmic bins in $v$ for every $N$. The entrance-density and cutoff estimators are
\begin{equation}
 q_N(v)=\frac{\ell}{n_0}N^{2/3}p_{N,v}(v),
 \quad
 G_N(v)=2\sqrt\pi v^{3/2}q_N(v).
\end{equation}
Only bins with at least 15 counts, $S\geq S_{\min}$, and
\begin{equation}
 v\geq \frac{c_{\mathrm{res}}}{E_{N_{\rm modes}-1}(\eta)}
\end{equation}
were retained. The final condition excludes the small-$v$ range unresolved by the truncated spectral sum. We propagated Poisson counting errors to $G_N$. The values used were $S_{\min}=30$ and $c_{\mathrm{res}}=8$. Because the fixed-seed edge mass decreases as $N^{-2/3}$, the density-level comparison is noisier than the transform and generalized-eigenvalue tests.

\end{document}